\documentclass[11pt,a4paper]{article}
\usepackage{jheppub}

\title{\boldmath Inadequacy of zero-width approximation for a light Higgs boson signal}
\author[a]{Nikolas Kauer}
\author[b]{and Giampiero Passarino}
\affiliation[a]{Department of Physics, Royal Holloway, University of London,\\ Egham TW20 0EX, United Kingdom}
\affiliation[b]{Dipartimento di Fisica Teorica, Universit\`a di Torino, Italy\\
           INFN, Sezione di Torino, Italy}
\emailAdd{n.kauer@rhul.ac.uk}
\emailAdd{giampiero@to.infn.it}

\abstract{
In the Higgs search at the LHC, a light Higgs boson ($115$\,GeV $\lesssim M_H \lesssim$ 130\,GeV) is
not excluded by experimental data. In this mass range, the width of the Standard Model Higgs 
boson is more than four orders of magnitude smaller than its mass. The zero-width 
approximation is hence expected to be an excellent approximation. We show that this is not 
always the case. The inclusion of off-shell contributions is essential to obtain an accurate 
Higgs signal normalisation at the $1\%$ precision level. For $gg\ (\to H) \to VV$, $V= W,Z$, 
${\cal O}(10\%)$ corrections occur due to an enhanced Higgs signal in the region $M_{VV} > 2\,M_V$,
where also sizable Higgs-continuum interference occurs. We discuss how experimental selection 
cuts can be used to exclude this region in search channels where the Higgs invariant mass cannot 
be reconstructed.
We note that the $H\to VV$ decay modes in weak boson fusion are similarly affected.
}


\arxivnumber{1206.4803}

\newcommand{\sla}[1]{\ifmmode%
  \setbox0=\hbox{$#1$}%
  \setbox1=\hbox to\wd0{\hss$/$\hss}\else%
  \setbox0=\hbox{#1}%
  \setbox1=\hbox to\wd0{\hss/\hss}\fi%
  #1\hskip-\wd0\box1 }
\newcommand{\calM}{{\cal M}}
\newcommand{\calO}{{\cal O}}

\newcommand{\ssF}{{\mathrm{F}}}

\newcommand{\sH}{\mathrm{H}}

\newcommand{\ssZ}{{\mathrm{Z}}}
\newcommand{\ssZZ}{{\mathrm{ZZ}}}
\newcommand{\bqas}{\begin{eqnarray*}}
\newcommand{\eqas}{\end{eqnarray*}}
\newcommand{\nl}{\nonumber\\}

\newcommand{\lpar}{\left(}                            
\newcommand{\rpar}{\right)}

\newcommand{\bq}{\begin{equation}}                    
\newcommand{\eq}{\end{equation}}
\newcommand{\bqa}{\arraycolsep 0.14em\begin{eqnarray}}
\newcommand{\eqa}{\end{eqnarray}}
\newcommand{\ba}[1]{\begin{array}{#1}}
\newcommand{\ea}{\end{array}}
\newcommand{\ben}{\begin{enumerate}}
\newcommand{\een}{\end{enumerate}}
\newcommand{\bei}{\begin{itemize}}
\newcommand{\eei}{\end{itemize}}
\newcommand{\eqn}[1]{Eq.\ (\ref{#1})}

\newcommand{\bmid}{\Bigr|}

\newcommand{\Bref}[1]{Ref.~\cite{#1}}

\newcommand{\UTeV}{{\,\text{TeV}}}
\newcommand{\UGeV}{{\,\text{GeV}}}
\newcommand{\UMeV}{{\,\text{MeV}}}
\newcommand{\Upb}{{\,\text{pb}}}

\newcommand{\PH}{H}
\newcommand{\PZ}{Z}
\newcommand{\PW}{W}
\newcommand{\PV}{V}
\newcommand{\Pg}{g}

\newcommand{\mBh}{{{\overline M}_{\PH}}}
\newcommand{\mBhs}{{{\overline M}^2_{\PH}}}

\newcommand{\cph}{{s_H}}

\newcommand{\muh}{{\mu_{\PH}}}
\newcommand{\muhs}{{\mu^2_{\PH}}}
\newcommand{\gh}{{\gamma_{\PH}}}

\newcommand{\GOL}{{\overline\Gamma}}

\providecommand\HTO{{\sc HTO}}

\begin{document}


\maketitle



\section{Introduction}

A key objective of current particle physics research is the experimental confirmation of a theoretically consistent description of elementary particle masses.  
In the Standard Model (SM), this is 
achieved through the Higgs mechanism \cite{Higgs:1964ia,Higgs:1964pj,Higgs:1966ev,Englert:1964et,Guralnik:1964eu}, which predicts the existence of 
one physical Higgs boson.  Searches for the SM Higgs boson have been carried out at the Large Electron Positron Collider (LEP)
and the Tevatron, which resulted in a lower Higgs mass bound of $114.4\,$GeV \cite{Barate:2003sz}, and the exclusion of $M_H\in[147,180]\,$GeV and $M_H\in[100,103]\,$GeV \cite{:2012zzl}, respectively.\footnote{All bounds and exclusion limits are at 95\% C.L.}  Higher sensitivity is attainable at 
the Large Hadron Collider (LHC), which was built as Higgs discovery machine.
The combined analysis of the CMS 2011 data of 4.6--$4.8\,$fb$^{-1}$ at $7\,$TeV
excludes $M_H\in[127,600]\,$GeV \cite{Chatrchyan:2012tx}.
Similarly, a recent ATLAS study excludes $M_H\in[111.4,116.6]\,$GeV, 
$M_H\in[119.4,122.1]\,$GeV and $M_H\in[129.2,541]\,$GeV \cite{:2012an}.
A light Higgs boson is therefore not 
excluded by experimental data.  
In fact, in a seminar on 4th July 2012 at CERN, ATLAS and CMS have presented 
evidence that 
a SM-like Higgs boson with $M_H\approx 125$--$126$ GeV has been observed at the 
$5\sigma$ level.
It is therefore important to examine the accuracy of theoretical 
predictions for Higgs production and decay at the LHC that are 
used in experimental analyses for light Higgs masses.

For light Higgs masses, 
the loop-induced gluon-fusion production ($gg\to H$) \cite{Georgi:1977gs} 
dominates.  Next-to-leading order (NLO) QCD corrections have been calculated in 
the heavy-top limit \cite{Dawson:1990zj} as well as with finite $t$ 
and $b$ mass effects \cite{Djouadi:1991tka,Graudenz:1992pv,Spira:1995rr}.  NLO corrections of 80--100\% at the LHC motivated 
the calculation of next-to-next-to-leading order (NNLO) QCD corrections in the 
heavy-top limit 
\cite{Harlander:2002wh,Anastasiou:2002yz,Ravindran:2003um} 
enhanced by soft-gluon resummation at next-to-next-to-leading logarithmic (NNLL) 
level \cite{Catani:2003zt,deFlorian:2011xf} and beyond \cite{Moch:2005ky,Laenen:2005uz,Idilbi:2005ni,Ravindran:2005vv,Ravindran:2006cg,Ahrens:2008nc}.
Fully differential calculations have been presented in Refs.\ 
\cite{Anastasiou:2004xq,Catani:2007vq}.
The accuracy of the $M_t\to \infty$ approximation at NNLO 
has been investigated in Refs.\ 
\cite{Marzani:2008az,Harlander:2009bw,Pak:2009bx,Harlander:2009mq,Pak:2009dg,Harlander:2009my}.\footnote{Scale, PDF, strong coupling and heavy-top-limit uncertainties have recently 
been reappraised in Ref.\ \cite{Baglio:2010ae}.}
In addition to higher-order QCD corrections, electroweak (EW) corrections 
have been computed up to two loops 
\cite{Djouadi:1994ge,Aglietti:2004nj,Degrassi:2004mx,Actis:2008ug,Actis:2008ts,Ahrens:2010rs,Keung:2009bs,Brein:2010xj} and found to be at the 1--5\% level.  
Mixed QCD-EW effects have also been calculated \cite{Anastasiou:2008tj}.
Refined calculations/updated cross sections for $gg\to H$ 
have been presented in Refs.\ 
\cite{deFlorian:2009hc,Dittmaier:2011ti,Anastasiou:2011pi,Anastasiou:2012hx,deFlorian:2012yg}.
Kinematic distributions and NNLO cross sections with experimental selection 
cuts have also been studied extensively for $gg\to H\to VV\to$ 4 leptons ($V=W,Z$)
\cite{Anastasiou:2007mz,Grazzini:2008tf} and all other important decay modes 
(see Ref.\ \cite{Dittmaier:2012vm} and references therein).
NLO EW corrections to $H\to VV \to$ 4 leptons have been calculated in Refs.\ \cite{Bredenstein:2006rh,Denner:2011vt}.

\begin{figure}[t]
\vspace{0.4cm}
\centering
\includegraphics[height=2.1cm, clip=true]{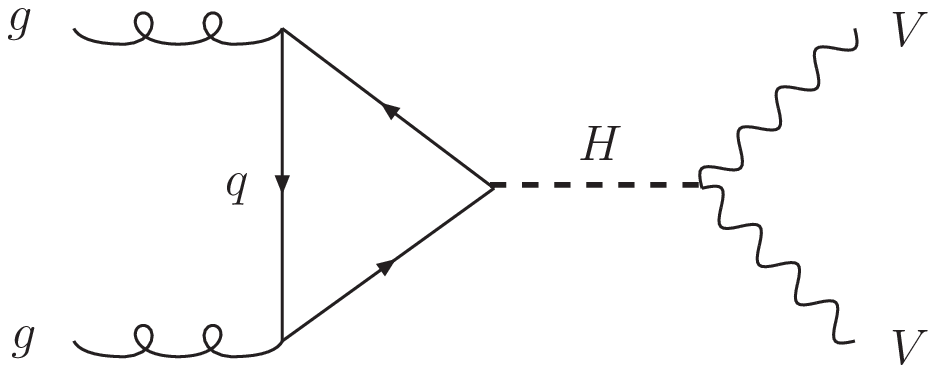}\hfil %
\includegraphics[height=2.1cm, clip=true]{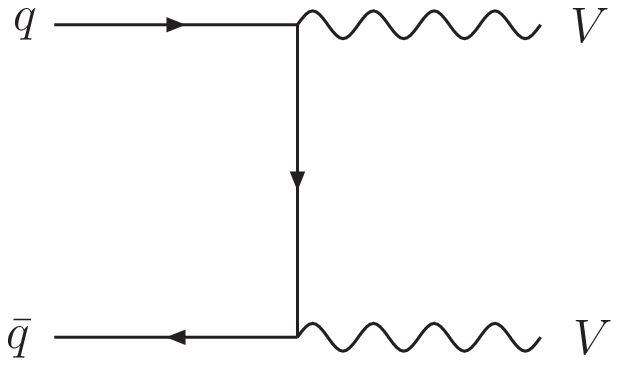}\hfil %
\includegraphics[height=2.1cm, clip=true]{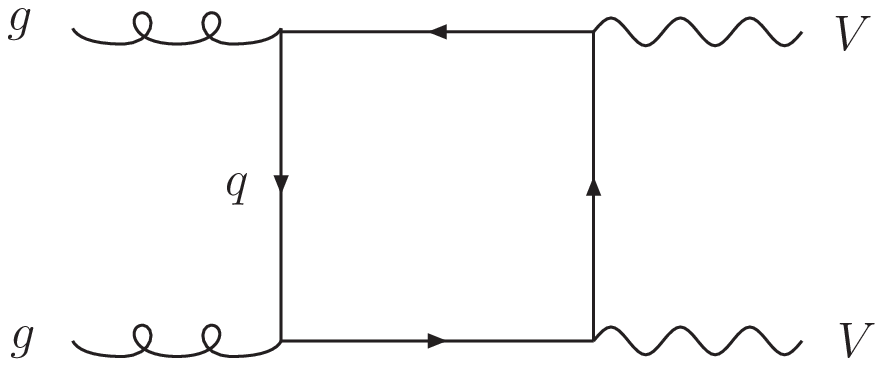}\\[.0cm]
\caption{\label{fig:graphs}
Representative Feynman graphs for the Higgs signal process (left) and the $q\bar{q}$- (center) and $gg$-initiated (right) continuum background processes at LO.}
\end{figure}

The proper theoretical description of the Higgs boson line shape is 
an essential ingredient for heavy Higgs searches and has been studied in 
detail in Ref.\ \cite{Goria:2011wa}.  A comparison of the zero-width 
approximation (ZWA, see below) and finite-width Higgs 
propagator schemes for inclusive Higgs production and decay can also be found 
in Refs.\ \cite{Anastasiou:2011pi,Buehler:2012zf,Anastasiou:2012hx}.\footnote{The accuracy of the 
ZWA in the context of beyond-the-SM physics has been studied 
in Refs.\ \cite{Berdine:2007uv,Kauer:2007zc,Kauer:2007nt,Uhlemann:2008pm}.}
In the light Higgs mass range the on-shell width of the SM Higgs boson is more 
than 
four orders of magnitude smaller than its mass, for instance 
$4.03\,$MeV for a  mass of $125\,$GeV.\footnote{Width computed with \HTO{}, see Section \protect\ref{sec:inclusive}.}
The ZWA a.k.a. narrow-width approximation,
which factorizes the Higgs cross section into on-shell production and on-shell
decay when $\Gamma_H$ approaches zero, is expected to be excellent well below 
the $WW$ and $ZZ$ thresholds with an error estimate of $\calO(\Gamma_H/M_H)$.
For Higgs production in gluon fusion, we show in Sections \ref{sec:inclusive} and 
\ref{sec:cuts} that this is not always the case.
For $gg\to H\to VV$, we find that the deviation between ZWA and off-shell 
results is particularly large. We therefore take into account the 
resonance-continuum interference (see Fig.~\ref{fig:graphs}, left and right), 
which was studied in Refs.\ \cite{Glover:1988fe,Glover:1988rg,Binoth:2006mf,Accomando:2007xc,Campbell:2011cu,Kauer:2012ma} 
and for related processes in Refs.\ \cite{Dixon:2003yb,Dixon:2008xc,Accomando:2011eu}. For studies of the continuum background (see Fig.~\ref{fig:graphs}, center and right), we refer the reader to Refs.\ \cite{Campbell:2011bn,Frederix:2011ss,Melia:2012zg,Agrawal:2012df} and references therein.

The paper is organised as follows: In Section \ref{sec:inclusive}, we briefly 
review 
the zero-width approximation and why it can be inadequate.  We then present 
and discuss inclusive results in ZWA and with off-shell effects for the processes 
$gg\to H\to$ all and $gg\to H\to ZZ$ with $M_H=125\,$GeV 
including Higgs-continuum interference effects.
In Section \ref{sec:cuts}, we extend our ZWA v.\ off-shell analysis by 
considering experimental Higgs search procedures, selection criteria and 
transverse mass observables for all $gg\to H\to VV\to$ 4 leptons search channels.
Higgs-continuum interference effects are again included.
Conclusions are given in Section \ref{sec:conclusions}.


\section{Inclusive analysis\label{sec:inclusive}}

In the SM, the common belief is that for a light Higgs boson the product of on-shell production 
cross-section (say in gluon-gluon fusion, $gg \to H$) and branching ratios reproduces the correct 
result to great accuracy. The expectation is based on the well-known result
\bqa
D_{\PH}(q^2) &=& \frac{1}{\lpar q^2 - M^2_{\PH}\rpar^2 + \Gamma^2_{\PH}\,M^2_{\PH}} =
\frac{\pi}{M_{\PH}\,\Gamma_{\PH}}\,\delta\lpar q^2 - M^2_{\PH}\rpar 
\nl
{}&+& PV\,\Bigl[ \frac{1}{\lpar q^2 - M^2_{\PH}\rpar^2}\Bigr] + \sum_{n=0}^N\,c_n(\alpha)\,
\delta_n\lpar q^2 - M^2_{\PH}\rpar 
\label{expBW}
\eqa
where $q^2$ is the virtuality of the Higgs boson, $M_{\PH}$ and $\Gamma_{\PH}$ are the on-shell 
Higgs mass and width and $PV$ denotes the principal value (understood as a distribution).
Furthermore, $\delta_n(x)$ is connected to the $n$th derivative of the delta-function by
$\delta_n(x)= (-1)^n/n\,!\,\delta^{(n)}(x)$ and the expansion is in terms of the coupling 
constant, up to a given order $N$.

In general, the ZWA can be applied to predict the probability for resonant scattering processes 
when the total decay width $\Gamma$ of the resonant particle is much smaller than its mass $M$.
Note that both concepts, on-shell mass and width, are ill-defined for an unstable particle and
should be replaced with the complex pole, which is a property of the $S\,$-matrix, gauge-parameter
independent to all orders of perturbation theory. Nevertheless, let us continue with our
qualitative argument: in the limit $\Gamma\to 0$, the mod-squared propagator
\bq
D(q^2)= \left[\left(q^2-M^2\right)^2+(M\Gamma)^2\right]^{-1}
\eq
with 4-momentum $q$ approaches the delta-function limit of \eqn{expBW}, i.e.
\bq
D(q^2) \sim K\,\delta(q^2-M^2),
\qquad  
K= \frac{\pi}{M\Gamma} = \int_{-\infty}^{+\infty} dq^2\,D(q^2).
\eq
The scattering cross-section $\sigma$ thus approximately decouples into on-shell 
production ($\sigma_p$) and decay as shown in Eqs.~(\ref{eq:ZWA1})--(\ref{eq:ZWA3}), 
where $s$ is the total 4-momentum squared, 
argument based on the scalar nature of the resonance. 
Based on the scales occurring in $D(q^2)$, the conventional error 
estimate is $\calO(\Gamma/M)$.  This will not be accurate when the $q^2$ dependence 
of $|\calM_p|^2$ or $|\calM_d|^2$ is strong enough to compete with the $q^2$ dependence of $D$.  
An interesting example is $gg\to H\to VV$, where $\sum|\calM_d(q^2)|^2\sim (q^2)^2$ above $2\,M_V$.
We note that similar effects have been observed for processes in SM extensions 
\cite{Berdine:2007uv,Kauer:2007zc,Kauer:2007nt,Uhlemann:2008pm}.
\begin{gather}
\label{eq:ZWA1}
\sigma = \frac{1}{2s}\left[\int_{q^2_\text{min}}^{q^2_\text{max}}
\frac{dq^2}{2\pi}\left(\int d\phi_p|\calM_p(q^2)|^2 
D(q^2)\, \int d\phi_d|\calM_d(q^2)|^2\right)\right]\\
\label{eq:ZWA2}
\sigma_\text{ZWA} = \frac{1}{2s}\left(\int d\phi_p|\calM_p(M^2)|^2 \right) 
\left(\int_{-\infty}^\infty   \frac{dq^2}{2\pi}\,D(q^2)
\right) \left( \int d\phi_d|\calM_d(M^2)|^2 \right)\\
\label{eq:ZWA3}
\sigma_\text{ZWA} = \frac{1}{2s}\left(\int d\phi_p|\calM_p|^2 \right) 
\frac{1}{2M\Gamma} \left( \int d\phi_d|\calM_d|^2 \right)\bigg|_{q^2=M^2}
\end{gather}
An important observation is that the Breit-Wigner distribution does not drop off nearly as fast 
as, for instance, a Gaussian. The relative contribution of the tail more than $n$ widths from 
the peak can be estimated as $1/(n\pi)$, because \cite{Kauer:2001sp}
\begin{equation}
\label{eq:BWtail}
\int_{(M-n\Gamma)^2}^{(M+n\Gamma)^2}\frac{dq^2}{2\pi}\frac{1}{(q^2-M^2)^2+(M\Gamma)^2}\approx \frac{1}{2M\Gamma}\left(1-\frac{1}{n\pi}\right) .
\end{equation}
Since the width of a light Higgs is so small, $n=1000$ corresponds to only a few
GeV, beyond which one would expect less than $0.04\%$ of the signal cross section.

A potential worry, addressed in this paper, is: to which level of accuracy does the ZWA 
approximate the full off-shell result, given that at $M_{\PH} = 125\UGeV$ the on-shell width (very
close to the imaginary part of the complex pole) is $4.03\UMeV$. When searching for the Higgs boson 
around $125\UGeV$ one should not care about the region $M_{\ssZZ} > 2\,M_{\PZ}$ but, due to 
limited statistics, theory predictions for the normalisation in $\bar{q}q - \Pg\Pg \to  \PZ\PZ$ 
are used over the entire spectrum in the $\PZ\PZ$ invariant mass. 

Therefore, the question is not to dispute that off-shell effects are depressed by a factor
$\Gamma_{\PH}/M_{\PH}$, as shown in \eqn{expBW}, but to move away from the peak in the
invariant mass distribution and look at the behavior of the distribution, no matter how small 
it is compared to the peak; is it really decreasing with $M_{\ssZZ}$? Is there a plateau? For how 
long is the plateau lasting? How does that affect the total cross-section if no cut is made?

In this section, we consider the signal (S) in the complex-pole scheme (CPS) of 
Refs.\ \cite{Goria:2011wa,Passarino:2010qk,Actis:2006rc}
\bq
\sigma_{\Pg \Pg \to \PZ \PZ}(S) =
\sigma_{\Pg \Pg \to \PH \to \PZ \PZ}(M_{\ssZZ}) = \frac{1}{\pi}\,
\sigma_{\Pg \Pg \to \PH}\lpar M_{\ssZZ}\rpar\,\frac{M^4_{\ssZZ}}{\bmid M^2_{\ssZZ} - \cph\bmid^2}\,
\frac{\Gamma_{\PH \to \PZ\PZ}\lpar M_{\ssZZ}\rpar}{M_{\ssZZ}},
\eq
\label{signal}
where $\cph$ is the Higgs complex pole, parametrized by $\cph = \muhs - i\,\muh\,\gh$.
Note that $\gh$ is not the on-shell width, although the numerical difference is tiny for low
values of $\muh$, as shown in Ref.\ \cite{Goria:2011wa}.

The production cross-section, $\sigma_{\Pg \Pg \to \PH}$, is computed
with NNLO QCD corrections (see Ref.\ \cite{Dittmaier:2012vm}) and NLO 
EW ones~\cite{Actis:2008ug}. The partial decay width of the off-shell Higgs boson 
of virtuality $M_{\ssZZ}$ ($\Gamma_{\PH \to \PZ\PZ}$), is computed at NLO with leading NNLO effects in the 
limit of large Higgs boson mass, see Ref.\ \cite{Bredenstein:2007ec}.
Numerical results in this section are obtained with the program \HTO{} (G.~Passarino, unpublished) 
that allows for the study of the Higgs boson lineshape, in gluon-gluon fusion, 
using complex poles. Our results refer to $\sqrt{s} = 8\UTeV$ and are based on the MSTW2008 PDF 
sets~\cite{Martin:2009iq}. They are implemented according to the OFFP scheme, see
Eq.\ (45) of \Bref{Goria:2011wa}. Furthermore, we set the renormalization and factorization scale to the Higgs virtuality.

\begin{figure}[t]
\vspace{0.4cm}
\centering
\includegraphics[width=0.7\textwidth, clip=true,angle=0]{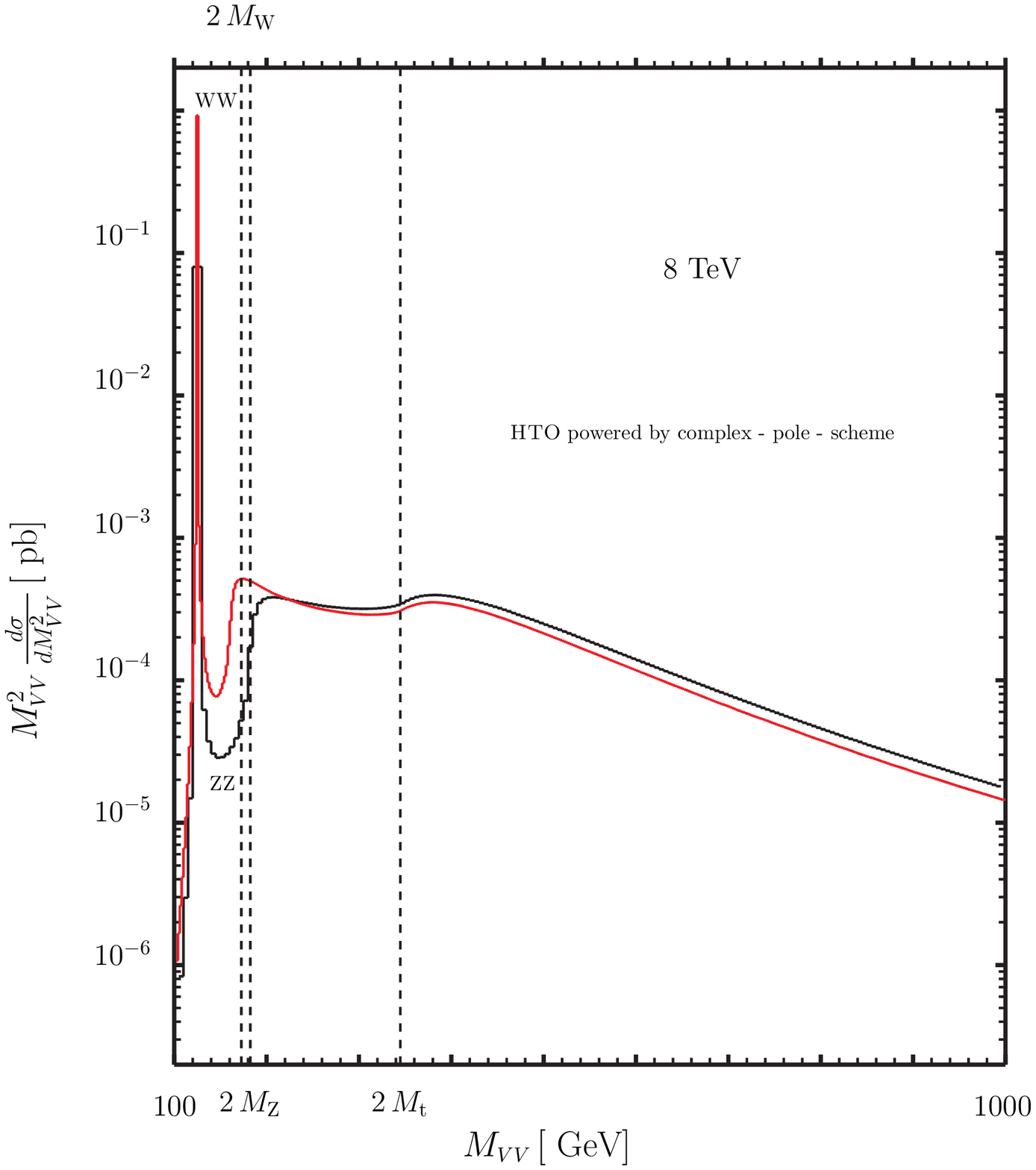}\\[.0cm]
\caption{\label{fig:HTO_1}
The NNLO $\PZ\PZ$ (black) and $\PW\PW$ (red) invariant mass distributions in $\Pg\Pg \to \PV\PV$ 
for $\muh = 125\UGeV$.}
\end{figure}

Away (but not too far away) from the narrow peak the propagator and the off-shell $\PH$ width 
behave like 
\bq
D_{\PH}\lpar M^2_{\ssZZ}\rpar \approx \frac{1}{\lpar M^2_{\ssZZ} - \muhs\rpar^2},
\qquad
\frac{\Gamma_{\PH \to \PZ\PZ}\lpar M_{\ssZ}\rpar}{M_{\ssZZ}} \sim G_{\ssF}\,M^2_{\ssZZ}
\eq
above threshold with a sharp increase just below it (it increases from $1.62\,\cdot\,10^{-2}\UGeV$
at $175\UGeV$ to $1.25\,\cdot\,10^{-1}\UGeV$ at $185\UGeV$).
Our result for the $\PV\PV$ ($\PV= \PW,\PZ$) invariant mass distribution is shown in 
Fig.\ \ref{fig:HTO_1}. 
It confirms that, above the peak, the distribution is decreasing until the effects of the 
$\PV\PV\,$ threshold become effective with a visible increase followed by a plateau, by another 
jump at the $t\bar{t}$-threshold, beyond which the signal distribution decreases almost linearly (on a logarithmic scale).
For $\Pg\Pg \to \PH \to \gamma \gamma$ the effect
is drastically reduced and confined to the region $M_{\gamma\gamma}$ between $157\,$GeV 
and $168\,$GeV, where the distribution is already five orders of magnitude below
the peak. 

What is the net effect on the total cross-section? We show it for $\PZ\PZ$ in Table \ref{tab:HTO_1} 
where the contribution above the $\PZ\PZ\,$-threshold amounts to $7.6\%$. We have checked that the 
effect does not depend on the propagator function, complex-pole propagator or
Breit-Wigner distribution. The size of the effect is related to the shape of
the distribution function. The complex-mass scheme can be translated into a more familiar 
language by introducing the Bar-scheme~\cite{Goria:2011wa}. Performing the well-known 
transformation
\bq
\mBhs = \muhs + \gamma^2_{\PH} \,,
\qquad
\mu_{\sH}\,{\GOL}_{\PH} = \mBh\,\gh.
\label{Bars}
\eq
a remarkable identity follows (defining the so-called Bar-scheme):
\bq
\frac{1}{M^2_{\ssZZ} - \cph} =
\Bigl( 1 + i\,\frac{{\GOL}_{\PH}}{\mBh}\Bigr)\,
\Bigr( M^2_{\ssZZ} - \mBhs + 
i\,\frac{{\GOL}_{\PH}}{\mBh}\,M^2_{\ssZZ} \Bigr)^{-1},
\label{barid}
\eq
showing that the complex-pole scheme is equivalent to introducing a running width in the 
propagator with parameters that are not the on-shell ones. Special attention goes to the
numerator in \eqn{barid} which is essential in providing the right asymptotic
behavior when $M_{\ssZZ} \to \infty$, as needed for cancellations with the rest of the
amplitude. Therefore, it is not advisable to use a naive, running-width Breit-Wigner
distribution or to use a propagator with $M^2_{\ssZZ} - M^2_{\PH} + i\,M_{\PH}\,
\Gamma_{\PH}(M^2_{\ssZZ})$.

\begin{table}[t]
\vspace{0.4cm}
\renewcommand{\arraystretch}{1.2}
\centering
\begin{tabular}{|c|ccc|}
\cline{2-4}
\multicolumn{1}{c|}{} & Tot[\Upb] & $M_{\ssZZ} > 2\,M_{\PZ}[\Upb]$ & R[\%] \\
\hline 
$\Pg\Pg \to \PH \to\,$ all  &  $19.146$  & $0.1525$ & $0.8$  \\
$\Pg\Pg \to \PH \to \PZ\PZ$ &  $0.5462$  & $0.0416$ & $7.6$  \\
\hline
\end{tabular}\\[.0cm]
\caption[]{\label{tab:HTO_1}{
Total cross-section for the processes $\Pg\Pg \to \PH \to \PZ\PZ$ and $\Pg\Pg \to \PH \to\,$ all;
the part of the cross-section coming from the region $M_{\ssZZ} > 2\,M_{\PZ}$ is explicitly 
shown, as well as the ratio.}}
\end{table}

In Table \ref{tab:HTO_2}, we present the invariant mass distribution integrated 
bin-by-bin.
%
\begin{table}[t]
\vspace{0.4cm}
\renewcommand{\arraystretch}{1.}
\centering
{\normalsize
\begin{tabular}{|ccccccc|}
\hline 
$100$--$125$ & $125$--$150$ & $150$--$175$ & $175$--$200$ & $200$--$225$ & $225$--$250$ & $250$--$275$ \\
\hline 
$0.252$ & $0.252$ & $0.195\cdot 10^{-3}$ & $0.177\cdot 10^{-2}$ &
$0.278\cdot 10^{-2}$ & $0.258\cdot 10^{-2}$ &
$0.240\cdot 10^{-2}$ \\
\hline
\end{tabular}}\\[.0cm]
\caption[]{\label{tab:HTO_2}{
Bin-by-bin integrated cross-section for the process $\Pg\Pg \to \PH \to \PZ\PZ$. The first row 
gives the bin in \UGeV, the second row gives the corresponding cross-section in \Upb.}}
\end{table}
%
If we take the ZWA value for the production cross-section at $8\UTeV$ and for 
$\muh= 125\UGeV$ ($19.146\Upb$) and use the branching ratio into $\PZ\PZ$ of 
$2.67\,\cdot\,10^{-2}$ we obtain a ZWA result of $0.5203\Upb$ with a $5\%$ difference 
w.r.t. the off-shell result, fully compatible with the $7.6\%$ effect coming from the high-energy 
side of the resonance.
In Table \ref{tab:HTO_1}, we also see that the effect is much less evident if we sum over 
all final states with a net effect of only $0.8\%$.
This agrees well with the deviation of 0.5\% between ZWA and 
fixed-width Breit-Wigner scheme (FWBW) given in Table 1 of 
Ref.\ \cite{Anastasiou:2011pi} for $M_H=120\,$GeV.
At $M_H=125\,$GeV, de Florian-Grazzini obtain a 0.3\%--0.4\% deviation 
between ZWA and CPS (or FWBW) with ``pure massless NNLO,''
i.e. without resummation, heavy quark effects and EW corrections,
and a slightly smaller deviation for the full calculation 
\cite{deFlorianPrivComm}.  For $gg\to H\to$ all, one can thus
expect deviations of ${\cal O}(1\%)$ depending on the particular 
implementation of the calculation.

\begin{figure}[t]
\vspace{0.4cm}
\centering
\includegraphics[width=0.7\textwidth,clip=true,angle=0]{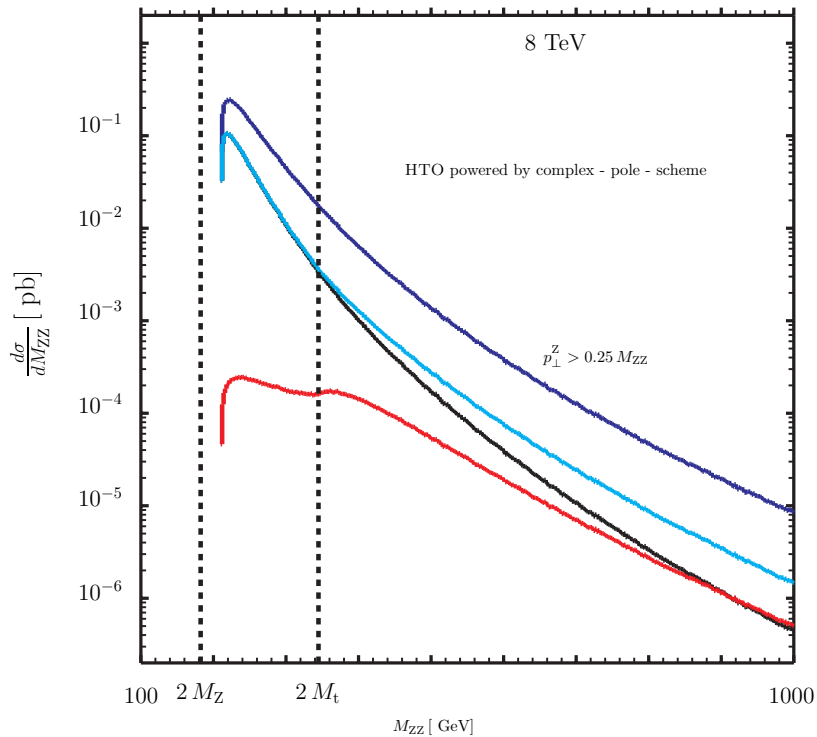}\\[.0cm]
\caption{\label{fig:HTO_2}
The LO $\PZ\PZ$ invariant mass distribution $\Pg\Pg \to \PZ\PZ$ for $\muh = 125\UGeV$. The 
black line is the total, the red line gives the signal while the cyan line gives signal plus
background; the blue line includes the $q\bar{q} \to \PZ\PZ$ contribution.}
\end{figure}

Of course, the signal per se is not a physical observable and one should always include
background and interference. In Fig.\ \ref{fig:HTO_2} we show the complete LO result 
for $gg \to ZZ$ calculated with {\sc HTO} with a cut of $0.25\,M_{\ssZZ}$ on the 
transverse momentum of the $\PZ$. The large destructive effects of the interference above the
resonant peak wash out the peculiar structure of the signal distribution. If one includes the 
region $M_{\ssZZ} > 2\,M_{\PZ}$ in the analysis then the conclusion is: interference effects 
are relevant also for the low Higgs mass region, at least for the $ZZ(WW)$ final 
state.
It is worth noting again that the discussed effect on the signal has nothing to do with 
$\Gamma_{\PH}/M_{\PH}$ effects; above the $\PZ\PZ\,$-threshold the distribution is higher than
expected (although tiny w.r.t. the narrow peak) and stays roughly constant up to the $t\bar{t}$-threshold
after which we observe an almost linear decrease. This is why the total cross-section is affected 
(in the $\PZ\PZ$ final state) at the $5\%$ level.

To conclude our inclusive analysis, we note that our findings are driven by
the interplay between the $q^2$-dependence of the Higgs propagator and 
the decay matrix element.  They should hence not only apply to 
Higgs production in gluon fusion, but also to Higgs production in weak boson fusion 
(WBF).
The enhancement for $H\to VV$ above $M_{VV}$ may even be stronger in 
WBF, because $\sigma(q\bar{q}\to q\bar{q}H)$ 
decreases less rapidly than $\sigma(gg \to H)$ with increasing Higgs 
invariant mass.\footnote{Preliminary results for inclusive WBF Higgs production 
reveal effects similar to $gg\to H\to$ all, yielding a deviation of 1\%
between ZWA and FWBW (but no difference between FWBW and CPS as expected 
for $M_H=125\,$GeV) \cite{RebuzziPrivComm,VBFTwiki}.}


\section{Analysis with experimental selection cuts \label{sec:cuts}}

In this section, we adopt the common selection cuts definition
between ATLAS 
and CMS for $H\to VV$ processes ($V=W,Z$) \cite{HWWcuts,HZZcuts}
and calculate parton-level $gg\ \to H\to VV\to$ leptons cross sections at LO using 
\textsf{gg2VV} \cite{gg2VV} based on Refs.\ \cite{Binoth:2005ua,Binoth:2006mf,Binoth:2008pr,Kauer:2012ma,Hahn:1998yk,Hahn:2000kx}, 
with Higgs in ZWA as well as off-shell including interference with continuum $VV$ production (where $\gamma^\ast$ contributions are also included).\footnote{%
All cross sections are evaluated with a $p_T(V) > 1$\,GeV cut.
This technical cut prevents numerical instabilities when evaluating the 
continuum amplitude.}%
All results are given for a single lepton flavour combination.  
No flavour summation is carried out for charged leptons or neutrinos.
As input parameters, we use the specification of the 
LHC Higgs Cross Section Working Group in App. A of Ref.\ \cite{Dittmaier:2011ti}
with NLO $\Gamma_V$ and $G_\mu$ scheme.  Finite top and bottom quark mass effects 
are included.  Lepton masses are neglected.
We consider the Higgs masses $125$\,GeV and $200$\,GeV
with $\Gamma_H = 0.004434$\,GeV and $1.428$\,GeV,
respectively.  The Higgs widths have been calculated with \textsc{HDECAY} \cite{Djouadi:1997yw}. The fixed-width prescription is used for Higgs and weak boson 
propagators.
The renormalisation and factorisation scales are set to $M_H/2$.  
The PDF set MSTW2008NNLO \cite{Martin:2009iq} with 3-loop running for 
$\alpha_s(\mu^2)$ and $\alpha_s(M_Z^2)=0.11707$ is used.
The CKM matrix is set to the unit matrix, which causes a 
negligible error \cite{Kauer:2012ma}.

The accuracy of the ZWA Higgs cross section and the impact of off-shell 
effects is assessed with the ratio
\begin{equation}
R_0=\frac{\sigma_{H,\text{ZWA}}}{\sigma_{H,\text{offshell}}} \;.
\end{equation}
To facilitate comparison with off-shell $M_{VV}$
distributions, we define the ZWA $M_{VV}$ distribution 
as suggested by Eq.\ (\ref{eq:ZWA2}):
\begin{equation}
\label{eq:ZWA_MVV}
\left(\frac{d\sigma}{dM_{VV}}\right)_{\!\text{ZWA}} =\ \sigma_{H,\text{ZWA}}\;\frac{M_H\Gamma_H}{\pi}\;\frac{2M_{VV}}{\left(M_{VV}^2-M_H^2\right)^2+(M_H\Gamma_H)^2} \;.
\end{equation}
Each signal process $gg\to H\to VV\to$ leptons (with amplitude ${\cal M}_H$) and corresponding 
continuum background process $gg\to VV\to$ leptons (with amplitude ${\cal M}_\text{cont}$) have identical initial and final states.  Hence interference occurs,
and the distinction between signal and background cross sections becomes blurred:
\begin{equation}
|{\cal M}_\text{VV}|^2 = |{\cal M}_H + {\cal M}_\text{cont}|^2 = |{\cal M}_H|^2 + |{\cal M}_\text{cont}|^2 + 2\,\mbox{Re}({\cal M}_H{\cal M}^\ast_\text{cont}) \;.
\end{equation}
We assess interference effects using a $(S+B)$-inspired interference measure,
\begin{equation}
R_1=\frac{\sigma(|{\cal M}_\text{VV}|^2)}{\sigma(|{\cal M}_H|^2 + |{\cal M}_\text{cont}|^2)} \;,
\end{equation}
and a $(S/\sqrt{B})$-inspired measure,
\begin{equation}
R_2=\frac{\sigma(|{\cal M}_H|^2+2\,\mbox{Re}({\cal M}_H{\cal M}^\ast_\text{cont}))}{\sigma(|{\cal M}_H|^2)} \;.
\end{equation}
In the following, charged leptons are denoted by $\ell$.


\subsection[\texorpdfstring{$gg\to H\to ZZ\to \ell\bar{\ell}\ell\bar{\ell}$ and $\ell\bar{\ell}\ell'\bar{\ell'}$ at $M_H=125$\,GeV}{gg -> H -> ZZ -> ll~ll~ and ll~l'l'~ at mH=125 GeV}]{\texorpdfstring{\boldmath $gg\to H\to ZZ\to \ell\bar{\ell}\ell\bar{\ell}$ and $\ell\bar{\ell}\ell'\bar{\ell'}$ at $M_H=125$\,GeV}{gg -> H -> ZZ -> ll~ll~ and ll~l'l'~ at mH=125 GeV}}

The same- and different-flavour 4-charged-lepton channels
have been analysed by ATLAS \cite{ATLAS:2012ac} and CMS 
\cite{Chatrchyan:2012dg} for Higgs masses in the range 110--600\,GeV. 
In these search channels, the invariant mass of the intermediate 
Higgs ($M_{H^\ast}\equiv M_{ZZ}$) can be reconstructed.
The $M_{ZZ}$ spectrum is hence used 
as the discriminant variable in the final stage of 
the analysis, and the test statistic is evaluated with a binned 
maximum-likelihood fit of signal and background models to the 
observed $M_{ZZ}$ distribution.
For light Higgs masses, the observed $M_{ZZ}$ distribution is
dominated by experimental resolution effects and
for example fitted as Gaussian with a standard deviation
of 2--2.5\,GeV (or similar bin sizes are used).
Since the width of a light SM 
Higgs boson is 2--3 orders of magnitude smaller, one would
expect that the ZWA is highly accurate.  According to 
Eq.~(\ref{eq:BWtail}), the constraints on $M_{ZZ}$ mentioned 
above introduce an error of order 0.1\%.  Invariant masses 
above $2\,M_Z$, where large deviations 
from the Breit-Wigner shape occur, are excluded by the experimental 
procedure.  Higgs-continuum interference effects are
negligible.
For illustration, we compute the Higgs cross section in ZWA 
and off-shell including continuum interference in the vicinity of $M_H$,
more precisely $|M_{ZZ}-M_H| < 1$\,GeV.
To take into account the detector acceptance, we require 
$p_{T\ell} > 5$\,GeV and $|\eta_\ell| < 2.5$.
Leptons are separated using $\Delta R_{\ell\ell} > 0.1$.
Following Ref.\ \cite{ATLAS:2012ac}, we apply the cuts 
76\,GeV$\,< M_{\ell\bar{\ell},12} < 106$\,GeV
and
$15\,$GeV$\,<M_{\ell\bar{\ell},34} < 115$\,GeV.
The invariant mass of the same-flavour, opposite-sign
lepton pair closest to $M_Z$ is denoted by
$M_{\ell\bar{\ell},12}$.
$M_{\ell\bar{\ell},34}$ denotes the invariant mass of
the remaining lepton pair.
The $\gamma^\ast$ singularity for vanishing virtuality
is excluded by requiring $M_{\ell\bar{\ell}} > 4$\,GeV.%
\footnote{This cut is induced by the phase space generation.}
The results are displayed in Table\ \ref{tab:2l2l_4l}.

\begin{table}[t]
\vspace{0.4cm}
\renewcommand{\arraystretch}{1.2}
\centering
{\small
\begin{tabular}{|l|cccc|c|cc|}
\cline{2-5} 
\multicolumn{1}{c|}{} & \multicolumn{4}{|c|}{$gg\ (\to H)\to ZZ \to 4\ell$ and $2\ell\,2\ell'$} & \multicolumn{3}{|c}{} \\
\cline{2-8} 
\multicolumn{1}{c|}{} & \multicolumn{4}{|c|}{$\sigma$ [fb], $pp$, $\sqrt{s} = 8$\,TeV, $M_H=125$\,GeV} & ZWA & \multicolumn{2}{c|}{interference} \\
\hline
\multicolumn{1}{|c|}{mode} & $H_\text{ZWA}$ & $H_\text{offshell}$ & cont & $|H_\text{ofs}$+cont$|^2$ & $R_0$ & $R_1$ & $R_2$ \\
\hline
$\ell\bar{\ell}\,\ell\bar{\ell}$ & 0.0748(2) & 0.0747(2) & 0.000437(3) & 0.0747(6) &1.002(3) & 0.994(8) & 0.994(8) \\
$\ell\bar{\ell}\,\ell'\bar{\ell'}$ & 0.1395(2) & 0.1393(2) & 0.000583(2) & 0.1400(3) & 1.002(2) & 1.001(2) & 1.001(2) \\
\hline
\end{tabular}}\\[.0cm]
\caption{\label{tab:2l2l_4l}
Cross sections for $gg\ (\to H)\to ZZ \to \ell\bar{\ell}\ell\bar{\ell}$ and $\ell\bar{\ell}\ell'\bar{\ell'}$ in 
$pp$ collisions at $\sqrt{s} = 8$\,TeV for $M_H=125$\,GeV and $\Gamma_H = 0.004434$\,GeV 
calculated at LO with \textsf{gg2VV}.  
The zero-width approximation (ZWA) and off-shell 
Higgs cross sections, the continuum cross section and the sum of
off-shell Higgs and continuum cross sections including interference are given.  
The accuracy of the ZWA and the impact of off-shell 
effects are assessed with 
$R_0=\sigma_{H,\text{ZWA}}/\sigma_{H,\text{offshell}}$.
Interference effects are illustrated through 
$R_1=\sigma(|{\cal M}_H + {\cal M}_\text{cont}|^2)/\sigma(|{\cal M}_H|^2 + |{\cal M}_\text{cont}|^2)$ 
and $R_2=\sigma(|{\cal M}_H|^2+2\,\mbox{Re}({\cal M}_H{\cal M}^\ast_\text{cont}))/\sigma(|{\cal M}_H|^2)$.  $\gamma^\ast$ contributions are included in 
${\cal M}_\text{cont}$.
Applied cuts: $|M_{ZZ}-M_H| < 1$\,GeV, $p_{T\ell} > 5$\,GeV, $|\eta_\ell| < 2.5$, 
$\Delta R_{\ell\ell} > 0.1$, 
76\,GeV$\,< M_{\ell\bar{\ell},12} < 106$\,GeV and 
$15\,$GeV$\,<M_{\ell\bar{\ell},34} < 115$\,GeV (see main text),
$M_{\ell\bar{\ell}} > 4$\,GeV.
Cross sections are given for a single lepton flavour combination.  
No flavour summation is carried out for charged leptons or neutrinos.
The integration error is given in brackets.
}
\end{table}


\subsection[\texorpdfstring{$gg\to H\to W^-W^+\to \ell\bar{\nu}_\ell \bar{\ell}\nu_\ell$ at $M_H=125$\,GeV}{}]{\texorpdfstring{\boldmath $gg\to H\to W^-W^+\to \ell\bar{\nu}_\ell \bar{\ell}\nu_\ell$ at $M_H=125$\,GeV}{}\label{sec:WW}}

The $WW\to2\ell\,2\nu$ search channel has been analysed by ATLAS \cite{ATLAS_lvlv}
and CMS \cite{Chatrchyan:2012ty} for Higgs masses in the range 110--600\,GeV.
We apply the standard cuts 
$p_{T\ell} > 20$\,GeV, $|\eta_\ell| < 2.5$, $\sla{p}_T > 30$\,GeV and
$M_{\ell\ell} > 12$\,GeV.
As Higgs search selection cuts, we apply the standard cuts and 
in addition $M_{\ell\ell} <$ 50\,GeV and $\Delta\phi_{\ell\ell} < 1.8$.
Since $M_{H^\ast}$ cannot be reconstructed, ATLAS and CMS also use transverse mass 
observables $M_T$ that aim at approximating $M_{H^\ast}$.
Ref.\ \cite{ATLAS_lvlv} uses the transverse mass definition%
\footnote{
In the absence of additional observed final state particles,
the expressions for $M_T$ simplify due to ${\sla{\bf{p}}}_T=-{\bf{p}}_{T,\ell\ell}$.
}
\begin{gather}
\label{eq:MT1}
M_{T1}=\sqrt{(M_{T,\ell\ell}+\sla{p}_{T})^2-({\bf{p}}_{T,\ell\ell}+{\sla{\bf{p}}}_T)^2}
\end{gather}
with
\begin{gather}
\label{eq:MTll}
M_{T,\ell\ell}=\sqrt{p_{T,\ell\ell}^2+M_{\ell\ell}^2}
\end{gather}
and applies a $0.75M_H < M_{T1} < M_H$ cut for $M_H=125$\,GeV.
Ref.\ \cite{Chatrchyan:2012ty} uses the transverse mass definition
\begin{gather}
\label{eq:MT2}
M_{T2}= \sqrt{2\,p_{T,\ell\ell}\:\sla{p}_T(1-\cos\Delta\phi_{\ell\ell,\text{miss}})}\,,
\end{gather}
where $\Delta\phi_{\ell\ell,\text{miss}}$ is the angle between
${\bf{p}}_{T,\ell\ell}$ and ${\sla{\bf{p}}}_T$, and applies a $80\,\text{\,GeV} < M_{T2} < M_H$ cut for $M_H=125$\,GeV.

\begin{figure}[t]
\vspace{0.4cm}
\centering
\includegraphics[width=0.7\textwidth, clip=true,angle=0]{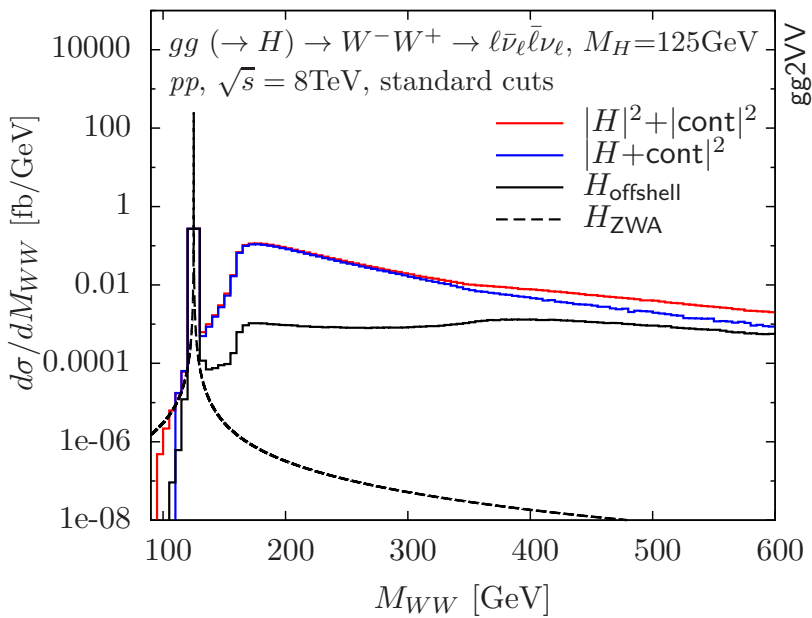}\\[.0cm]
\caption{\label{fig:lvlv_MWW_l}
$M_{WW}$ distributions for 
$gg\ (\to H)\to W^-W^+\to \ell\bar{\nu}_\ell \bar{\ell}\nu_\ell$ 
in $pp$ collisions at $\sqrt{s} = 8$\,TeV 
for $M_H=125$\,GeV and $\Gamma_H = 0.004434$\,GeV 
calculated at LO with \textsf{gg2VV}.
The ZWA distribution (black, dashed) as defined in Eq.~(\ref{eq:ZWA_MVV})
in the main text, the off-shell 
Higgs distribution (black, solid), the 
$d\sigma(|{\cal M}_H + {\cal M}_\text{cont}|^2)/dM_{WW}$ 
distribution (blue) and the 
$d\sigma(|{\cal M}_H|^2 + |{\cal M}_\text{cont}|^2)/dM_{WW}$ 
distribution (red) are shown.
Standard cuts are applied: 
$p_{T\ell} > 20$\,GeV, $|\eta_\ell| < 2.5$, $\sla{p}_T > 30$\,GeV, 
$M_{\ell\ell} >$ 12\,GeV.
Differential cross sections for a single lepton flavour combination are displayed. 
No flavour summation is carried out for charged leptons or neutrinos.
}
\end{figure}

\begin{figure}[t]
\vspace{0.4cm}
\centering
\includegraphics[width=0.7\textwidth, clip=true,angle=0]{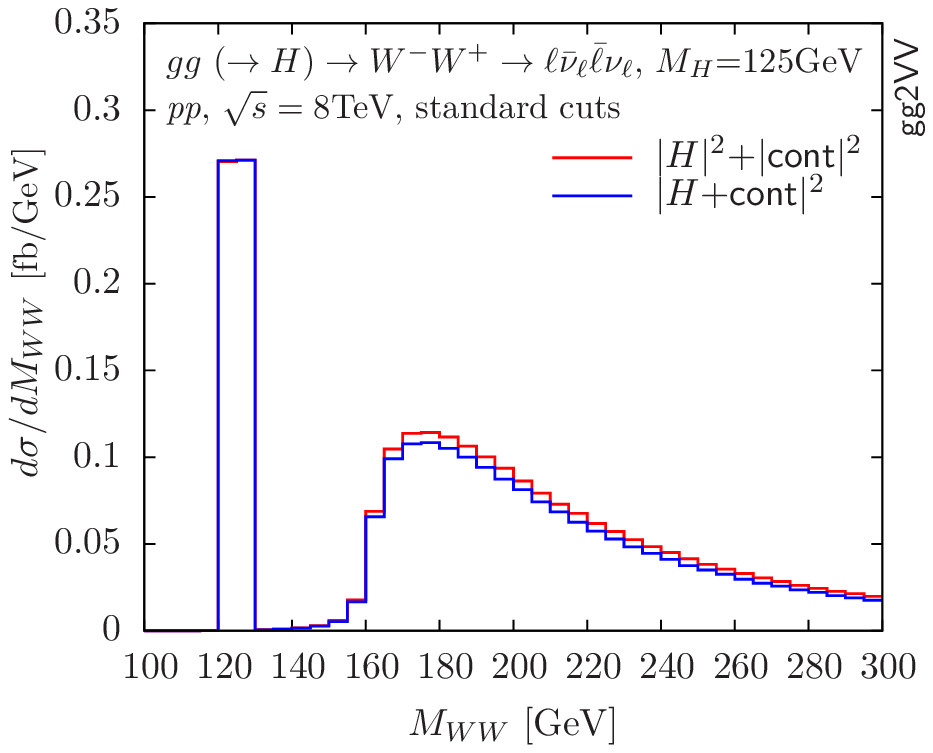}\\[.0cm]
\caption{\label{fig:lvlv_MWW_m}
$M_{WW}$ distributions for 
$gg\ (\to H)\to W^-W^+\to \ell\bar{\nu}_\ell \bar{\ell}\nu_\ell$ 
for $M_H=125$\,GeV.  Interference effects in the region of the
Higgs resonance and the $W$-pair threshold are shown.
Details as in Fig.\ \protect\ref{fig:lvlv_MWW_l}.
}
\end{figure}

\begin{figure}[t]
\vspace{0.4cm}
\centering
\includegraphics[width=0.7\textwidth, clip=true,angle=0]{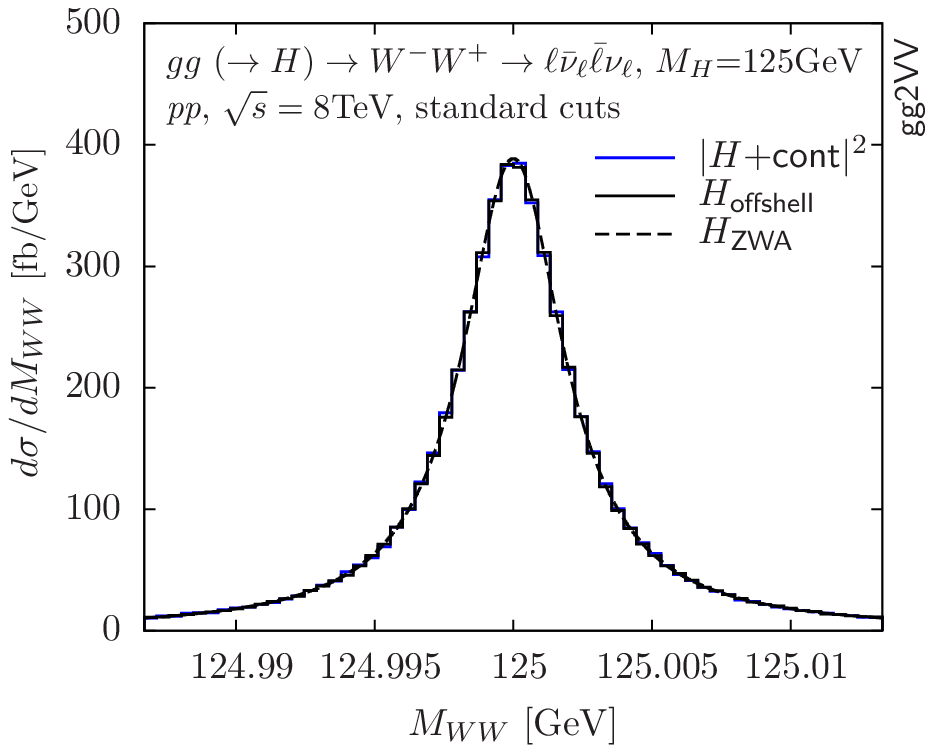}\\[.0cm]
\caption{\label{fig:lvlv_MWW_s}
$M_{WW}$ distributions for 
$gg\ (\to H)\to W^-W^+\to \ell\bar{\nu}_\ell \bar{\ell}\nu_\ell$ 
for $M_H=125$\,GeV.  Off-shell and interference effects in the vicinity of the
Higgs resonance are shown.
Details as in Fig.\ \protect\ref{fig:lvlv_MWW_l}.
}
\end{figure}

\begin{figure}[t]
\vspace{0.4cm}
\centering
\includegraphics[width=0.7\textwidth, clip=true,angle=0]{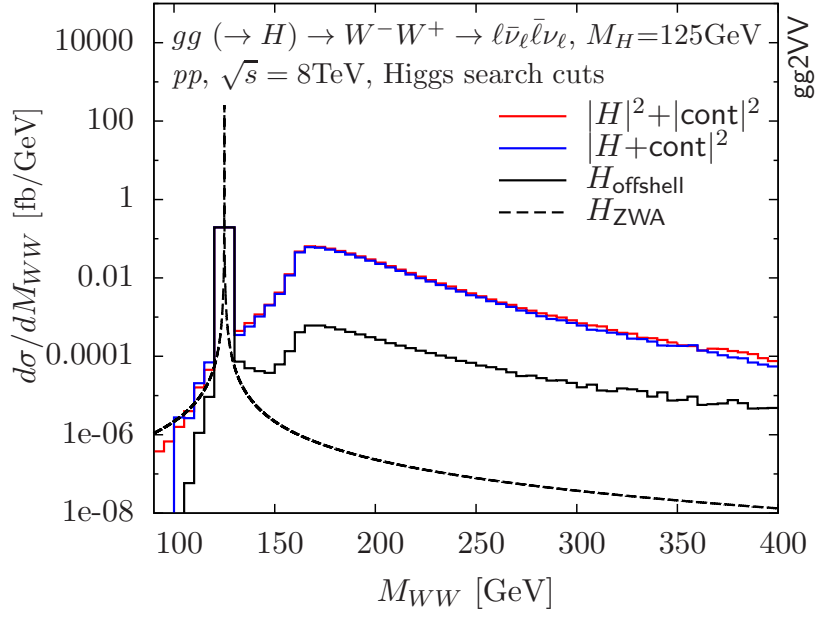}\\[.0cm]
\caption{\label{fig:lvlv_MWW_higgscuts}
$M_{WW}$ distributions for 
$gg\ (\to H)\to W^-W^+\to \ell\bar{\nu}_\ell \bar{\ell}\nu_\ell$ 
for $M_H=125$\,GeV.
Higgs search cuts are applied: 
$p_{T\ell} > 20$\,GeV, $|\eta_\ell| < 2.5$, $\sla{p}_T > 30$\,GeV, 
12\,GeV $<M_{\ell\ell} <$ 50\,GeV, $\Delta\phi_{\ell\ell} < 1.8$.
Other details as in Fig.\ \protect\ref{fig:lvlv_MWW_l}.
}
\end{figure}

Cross sections are presented in Table 
\ref{tab:lvlv}.  
When standard cuts are applied, the phase space region where $M_{WW}>160$\,GeV,
or equivalently $M_{WW} > M_H + 7000\Gamma_H$,
contributes 16\% to the off-shell Higgs cross section.  The error of the 
ZWA exceeds 15\%, and interference effects are of $\calO(10\%)$.  
Figs.\ \ref{fig:lvlv_MWW_l} and \ref{fig:lvlv_MWW_m} illustrate that 
the region with $M_{WW} > 2\,M_W$ is responsible for the
inaccuracy of the ZWA as well as the unexpectedly large interference effects,
in agreement with our discussion in Section \ref{sec:inclusive}.
Fig.\ \ref{fig:lvlv_MWW_s} demonstrates that finite-width effects and
Higgs-continuum interference are negligible in the resonance region, 
i.e.\ $|M_{WW} - M_H|\lesssim \Gamma_H$, for a Higgs mass of 125\,GeV.
The Higgs search selection has additional cuts, in particular an upper 
bound on the invariant mass of the observed dilepton system, which 
significantly reduce the contribution from the region with $M_{WW} \gg 2\,M_W$,
as seen in Fig.\ \ref{fig:lvlv_MWW_higgscuts}.  The result is a substantial 
mitigation of the finite-width and interference effects as seen in Table 
\ref{tab:lvlv}.

\begin{figure}[t]
\vspace{0.4cm}
\centering
\includegraphics[width=0.7\textwidth, clip=true,angle=0]{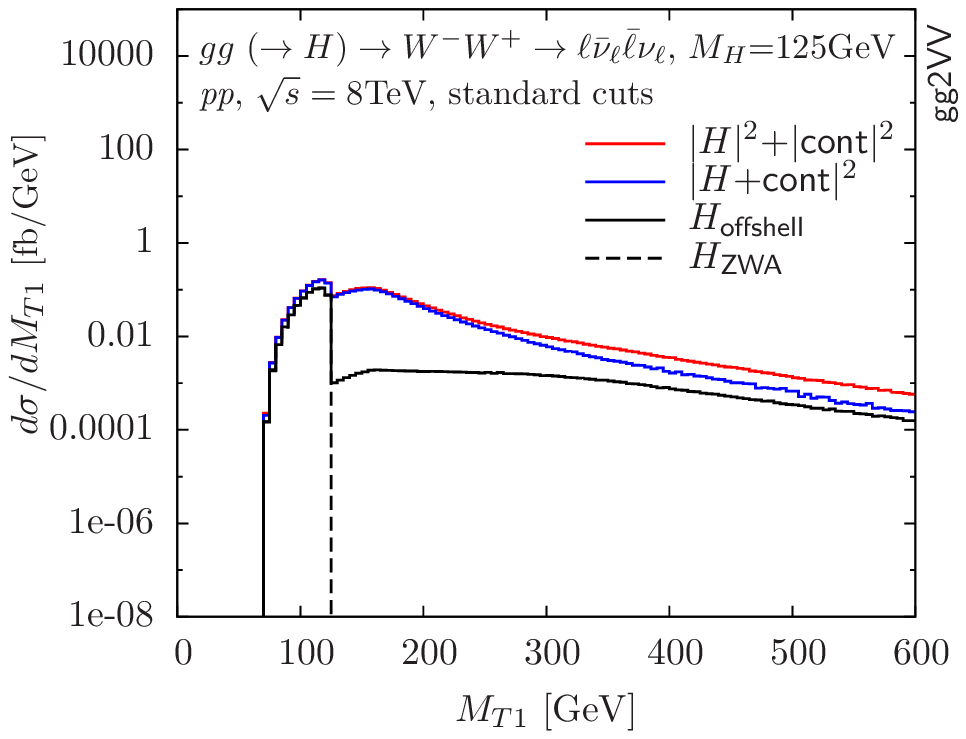}\\[.0cm]
\caption{\label{fig:lvlv_MT1_l}
Transverse mass distributions for 
$gg\ (\to H)\to W^-W^+\to \ell\bar{\nu}_\ell \bar{\ell}\nu_\ell$ 
for $M_H=125$\,GeV.  $M_{T1}$ is defined in Eq.\ (\ref{eq:MT1}) in the main text.
Other details as in Fig.\ \protect\ref{fig:lvlv_MWW_l}.
}
\end{figure}

\begin{figure}[t]
\vspace{0.4cm}
\centering
\includegraphics[width=0.7\textwidth, clip=true,angle=0]{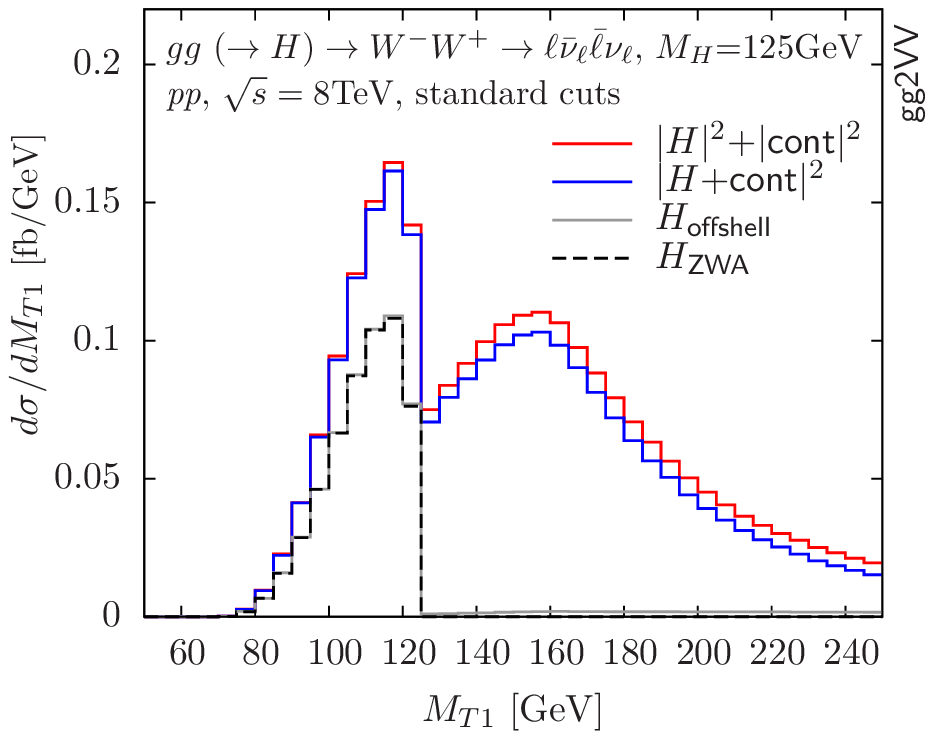}\\[.0cm]
\caption{\label{fig:lvlv_MT1_m}
$M_{T1}$ distributions for 
$gg\ (\to H)\to W^-W^+\to \ell\bar{\nu}_\ell \bar{\ell}\nu_\ell$ 
for $M_H=125$\,GeV.
Off-shell and interference effects in the region of the
Higgs resonance and the $W$-pair threshold are shown.
$M_{T1}$ is defined in Eq.\ (\ref{eq:MT1}) in the main text.
Other details as in Fig.\ \protect\ref{fig:lvlv_MWW_l}.
}
\end{figure}

\begin{figure}[t]
\vspace{0.4cm}
\centering
\includegraphics[width=0.7\textwidth, clip=true,angle=0]{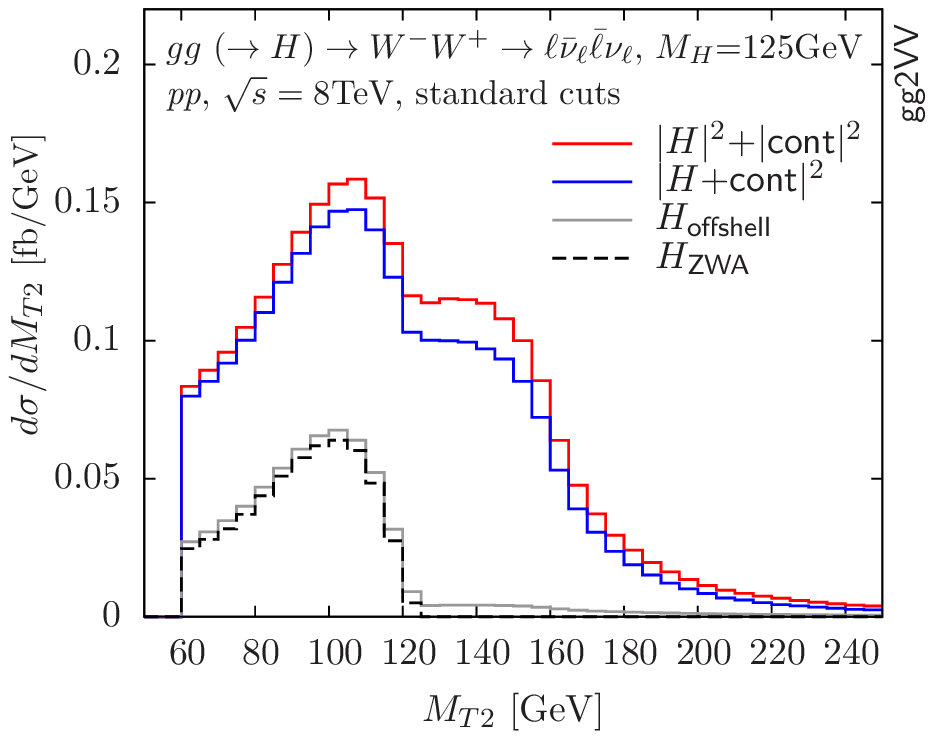}\\[.0cm]
\caption{\label{fig:lvlv_MT2_m}
$M_{T2}$ distributions for 
$gg\ (\to H)\to W^-W^+\to \ell\bar{\nu}_\ell \bar{\ell}\nu_\ell$ 
for $M_H=125$\,GeV.
Off-shell and interference effects in the region of the
Higgs resonance and the $W$-pair threshold are shown.
$M_{T2}$ is defined in Eq.\ (\ref{eq:MT2}) in the main text.
Other details as in Fig.\ \protect\ref{fig:lvlv_MWW_l}.
}
\end{figure}

We now consider the impact of cuts on transverse mass observables, which
are designed to have the physical mass of the decaying parent particle 
(the invariant mass in the off-shell case) as upper bound \cite{Barr:2009mx}.%
\footnote{We note that $M_{T1}$ is referred to as $M_T^\text{true}$ in Ref.\ \cite{Barr:2009mx}.}
For the process considered here, this is shown in Fig.\ \ref{fig:lvlv_MT1_l}.
Evidently, imposing a cut $M_T < M_H$ is an effective means to 
suppress interference effects.  This was first noticed and studied for the 
$M_{T1}$ variable in Ref.\ \cite{Campbell:2011cu}.
In Table \ref{tab:lvlv}, one can see that both, $M_{T1}$ and $M_{T2}$,
are suitable transverse mass variables, with $M_{T1}$ being slightly
more effective. 
This is also borne out by the transverse mass distributions 
in Figs.\ \ref{fig:lvlv_MT1_m} and \ref{fig:lvlv_MT2_m}.
With regard to the ZWA, Table \ref{tab:lvlv} shows that the application of the
$M_{T1}$ or $M_{T2}$ cut reduces the ZWA error to the sub-percent level.

\begin{table}[t]
\vspace{0.4cm}
\renewcommand{\arraystretch}{1.2}
\centering
{\scriptsize
\begin{tabular}{|c|cccc|c|cc|}
\cline{2-5} 
\multicolumn{1}{c|}{} & \multicolumn{4}{|c|}{$gg\ (\to H)\to W^-W^+\to \ell\bar{\nu}_\ell \bar{\ell}\nu_\ell$} & \multicolumn{2}{|c}{} \\
\cline{2-8} 
\multicolumn{1}{c|}{} & \multicolumn{4}{|c|}{$\sigma$ [fb], $pp$, $\sqrt{s} = 8$\,TeV, $M_H=125$\,GeV} & ZWA & \multicolumn{2}{c|}{interference} \\
\hline
\multicolumn{1}{|c|}{selection cuts} & $H_\text{ZWA}$ & $H_\text{offshell}$ & cont & $|H_\text{ofs}$+cont$|^2$ & $R_0$ & $R_1$ & $R_2$ \\
\hline
standard cuts & 2.707(3) & 3.225(3) & 10.493(5) & 12.241(8) & 0.839(2) & 0.8923(7) & 0.542(3) \\
Higgs search cuts & 1.950(1) & 1.980(1) & 2.705(2) & 4.497(3) & 0.9850(7) & 0.9599(7) & 0.905(2) \\
\hline
$0.75M_H < M_{T1} < M_H$ & 1.7726(9) & 1.779(1) & 0.6443(9) & 2.383(2) & 0.9966(8) & 0.983(1) & 0.977(2) \\
$80\,\text{\,GeV} < M_{T2} < M_H$ & 1.7843(9) & 1.794(1) & 0.955(1) & 2.687(3) & 0.9944(8) & 0.977(1) & 0.965(2) \\
\hline
\end{tabular}}\\[.0cm]
\caption{\label{tab:lvlv}
Cross sections for $gg\ (\to H)\to W^-W^+\to \ell\bar{\nu}_\ell \bar{\ell}\nu_\ell$ 
for $M_H=125$\,GeV with standard cuts, Higgs search cuts and additional transverse mass cut (either on $M_{T1}$ or $M_{T2}$).  
Standard cuts: $p_{T\ell} > 20$\,GeV, $|\eta_\ell| < 2.5$, $\sla{p}_T > 30$\,GeV, 
$M_{\ell\ell} >$ 12\,GeV.  Higgs search cuts: standard cuts and 
$M_{\ell\ell} <$ 50\,GeV, $\Delta\phi_{\ell\ell} < 1.8$.
$M_{T1}$ and $M_{T2}$ are defined in Eqs.\ (\ref{eq:MT1}) and 
(\ref{eq:MT2}) in the main text.
Other details as in Table\ \protect\ref{tab:2l2l_4l}.
}
\end{table}

\subsection[\texorpdfstring{$gg\to H\to ZZ \to \ell\bar{\ell}\nu_\ell\bar{\nu}_\ell$ at $M_H=200$\,GeV}{}]{\texorpdfstring{\boldmath $gg\to H\to ZZ \to \ell\bar{\ell}\nu_\ell\bar{\nu}_\ell$ at $M_H=200$\,GeV\label{sec:llvv_200}}{}}

The $ZZ\to2\ell\,2\nu$ search channel has been analysed by ATLAS \cite{ATLAS_2l2v}
and CMS \cite{Chatrchyan:2012ft} for Higgs masses in the range 200--600\,GeV. 
In this section we focus on the lowest studied Higgs mass of
$200$\,GeV with $\Gamma_H/M_H=0.7\%$.  Note that $M_H$ is slightly 
above the $Z$ pair production threshold.  A clean separation of
the Higgs resonance and the region with large continuum background
is thus not possible.
We apply the Higgs search cuts 
$p_{T\ell} > 20$\,GeV, $|\eta_\ell| < 2.5$, $76$\,GeV $< M_{\ell\ell} < 106$\,GeV,
$\sla{p}_T > 10$\,GeV and $\Delta\phi_{\ell\ell} > 1$.
Refs.\ \cite{ATLAS_2l2v} and \cite{Chatrchyan:2012ft} use a
transverse mass distribution as final discriminant in searching for 
an excess of data over the SM background expectation.
Ref.\ \cite{Chatrchyan:2012ft} employs the transverse mass 
variable first proposed in Ref.\ \cite{Rainwater:1999sd}
for the weak boson fusion $H\to WW$ channel:
\begin{gather}
\label{eq:MT3}
M_{T3}=\sqrt{ \left(M_{T,\ell\ell}+\sla{M}_{T}\right)^2-({\bf{p}}_{T,\ell\ell}+{\sla{\bf{p}}}_T)^2 }
\end{gather}
with $M_{T,\ell\ell}$ defined in Eq.\ (\ref{eq:MTll}) and
\begin{gather}
\label{eq:MTmiss}
\sla{M}_{T}=\sqrt{\sla{p}_{T}^2+M_{\ell\ell}^2}
\end{gather}
Note that $M_{T3}$, unlike $M_{T1}$ and $M_{T2}$, does 
not have a kinematic edge at $M_{H^\ast}$.  The variable used 
in Ref.\ \cite{ATLAS_2l2v} is obtained by replacing
$M_{\ell\ell}$ with $M_Z$ in the definition of $M_{T3}$, 
which causes only minor differences for $M_H>2M_Z$.
No $M_T$ cut is applied in the analyses.

\begin{figure}[t]
\vspace{0.4cm}
\centering
\includegraphics[width=0.7\textwidth, clip=true,angle=0]{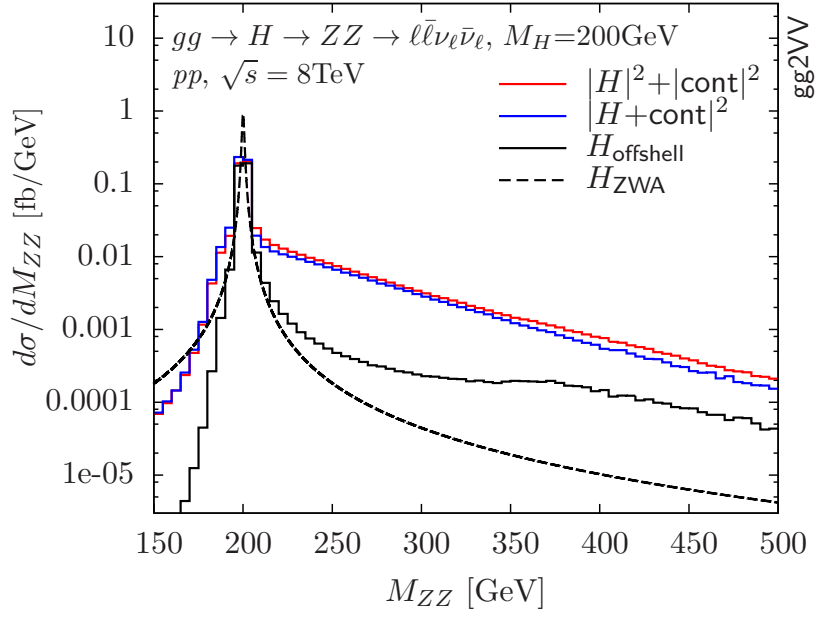}\\[.0cm]
\caption{\label{fig:2l2v_200_MZZ_l}
$M_{ZZ}$ distributions for $gg\to H\to ZZ\to \ell\bar{\ell}\nu_\ell\bar{\nu}_\ell$ 
for $M_H=200$\,GeV and $\Gamma_H = 1.428$\,GeV.
Applied cuts: $p_{T\ell} > 20$\,GeV, $|\eta_\ell| < 2.5$, 
$76$\,GeV $< M_{\ell\ell} < 106$\,GeV,
$\sla{p}_T > 10$\,GeV, $\Delta\phi_{\ell\ell} > 1$.
Other details as in Fig.\ \protect\ref{fig:lvlv_MWW_l}.
}
\end{figure}

\begin{figure}[t]
\vspace{0.4cm}
\centering
\includegraphics[width=0.7\textwidth, clip=true,angle=0]{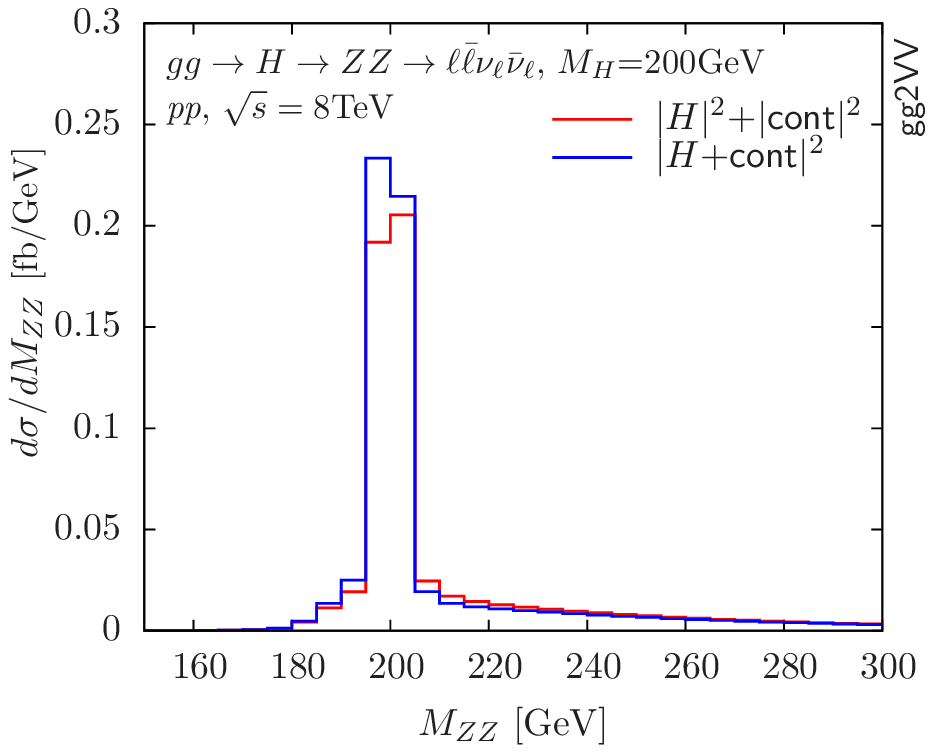}\\[.0cm]
\caption{\label{fig:2l2v_200_MZZ_m}
$M_{ZZ}$ distributions for $gg\to H\to ZZ\to \ell\bar{\ell}\nu_\ell\bar{\nu}_\ell$ 
for $M_H=200$\,GeV.  Interference effects in the region of the
Higgs resonance and the $Z$-pair threshold are shown.
Details as in Fig.\ \protect\ref{fig:2l2v_200_MZZ_l}.
}
\end{figure}

\begin{figure}[t]
\vspace{0.4cm}
\centering
\includegraphics[width=0.7\textwidth, clip=true,angle=0]{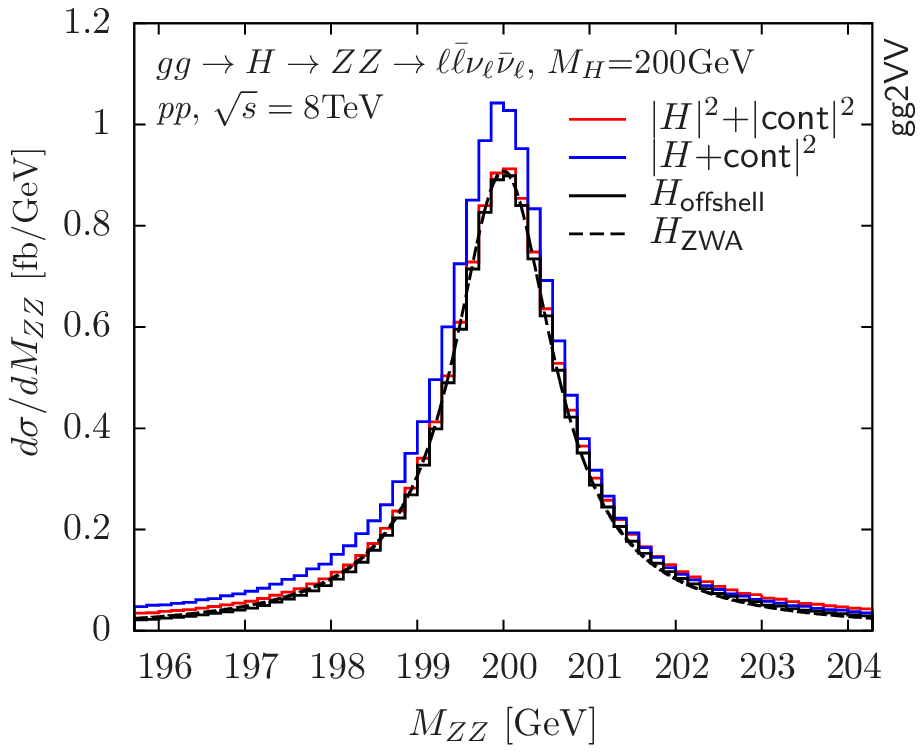}\\[.0cm]
\caption{\label{fig:2l2v_200_MZZ_s}
$M_{ZZ}$ distributions for $gg\to H\to ZZ\to \ell\bar{\ell}\nu_\ell\bar{\nu}_\ell$ 
for $M_H=200$\,GeV.
Off-shell and interference effects in the vicinity of the
Higgs resonance are shown.
Details as in Fig.\ \protect\ref{fig:2l2v_200_MZZ_l}.
}
\end{figure}

\begin{figure}[t]
\vspace{0.4cm}
\centering
\includegraphics[width=0.7\textwidth, clip=true,angle=0]{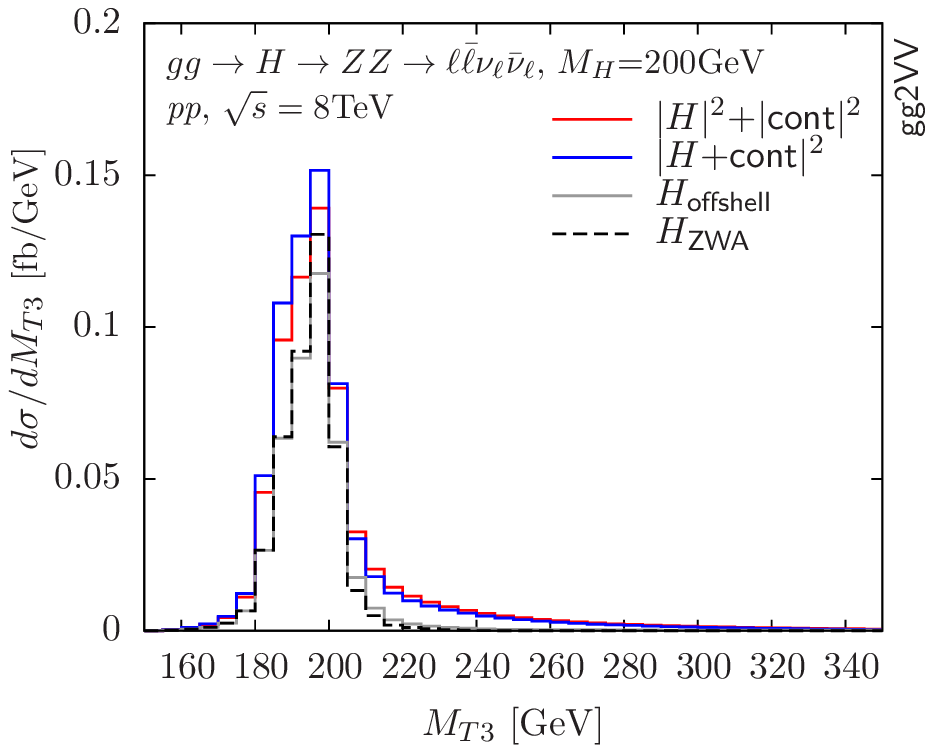}\\[.0cm]
\caption{\label{fig:2l2v_200_MT3_m}
Transverse mass distributions  
for $gg\to H\to ZZ\to \ell\bar{\ell}\nu_\ell\bar{\nu}_\ell$ for $M_H=200$\,GeV.
$M_{T3}$ is defined in Eq.\ (\ref{eq:MT3}) in the main text.
Other details as in Fig.\ \protect\ref{fig:2l2v_200_MZZ_l}.
}
\end{figure}

In Table \ref{tab:2l2v_200}, cross section results are given.
The ZWA is accurate at the percent level.  Fig.\ \ref{fig:2l2v_200_MZZ_l} reveals
that the off-shell enhancement of the high $M_{H^\ast}$ tail is moderate.
Higgs-continuum interference is constructive and of $\calO(5\%)$.  
Significant interference occurs in the vicinity of the Higgs resonance
as shown in Figs.\ \ref{fig:2l2v_200_MZZ_m} and \ref{fig:2l2v_200_MZZ_s}.
$ZZ$ interference effects are comparable to $WW$ interference 
effects for similar Higgs masses \cite{Binoth:2006mf,Campbell:2011cu}.
The $M_{T3}$ distributions displayed in Fig.\ \ref{fig:2l2v_200_MT3_m} 
show that sizable ZWA deviations occur at the differential level.

\begin{table}[t]
\vspace{0.4cm}
\renewcommand{\arraystretch}{1.2}
\centering
{\small
\begin{tabular}{|cccc|c|cc|}
\cline{1-4} 
\multicolumn{4}{|c|}{$gg\ (\to H)\to ZZ \to \ell\bar{\ell}\nu_\ell\bar{\nu}_\ell$} & \multicolumn{2}{|c}{} \\
\hline
\multicolumn{4}{|c|}{$\sigma$ [fb], $pp$, $\sqrt{s} = 8$\,TeV, $M_H=200$\,GeV} & ZWA & \multicolumn{2}{c|}{interference} \\
\hline
$H_\text{ZWA}$ & $H_\text{offshell}$ & cont & $|H_\text{ofs}$+cont$|^2$ & $R_0$ & $R_1$ & $R_2$ \\
\hline
2.0357(8) & 2.0608(9) & 1.1888(6) & 3.380(2) & 0.9878(6) & 1.0400(7) & 1.063(1) \\
\hline
\end{tabular}}\\[.0cm]
\caption{\label{tab:2l2v_200}
Cross sections for $gg\ (\to H)\to ZZ \to \ell\bar{\ell}\nu_\ell\bar{\nu}_\ell$ for $M_H=200$\,GeV and $\Gamma_H = 1.428$\,GeV.
Applied cuts: $p_{T\ell} > 20$\,GeV, $|\eta_\ell| < 2.5$, 
$76$\,GeV $< M_{\ell\ell} < 106$\,GeV,
$\sla{p}_T > 10$\,GeV, $\Delta\phi_{\ell\ell} > 1$.
Other details as in Table \protect\ref{tab:2l2l_4l}.
}
\end{table}

\subsection[\texorpdfstring{$gg\to H\to ZZ \to \ell\bar{\ell}\nu_\ell\bar{\nu}_\ell$ at $M_H=125$\,GeV}{}]{\texorpdfstring{\boldmath $gg\to H\to ZZ \to \ell\bar{\ell}\nu_\ell\bar{\nu}_\ell$ at $M_H=125$\,GeV}{}\label{sec:ZZ_125}}

\begin{figure}[t]
\vspace{0.4cm}
\centering
\includegraphics[width=0.7\textwidth, clip=true,angle=0]{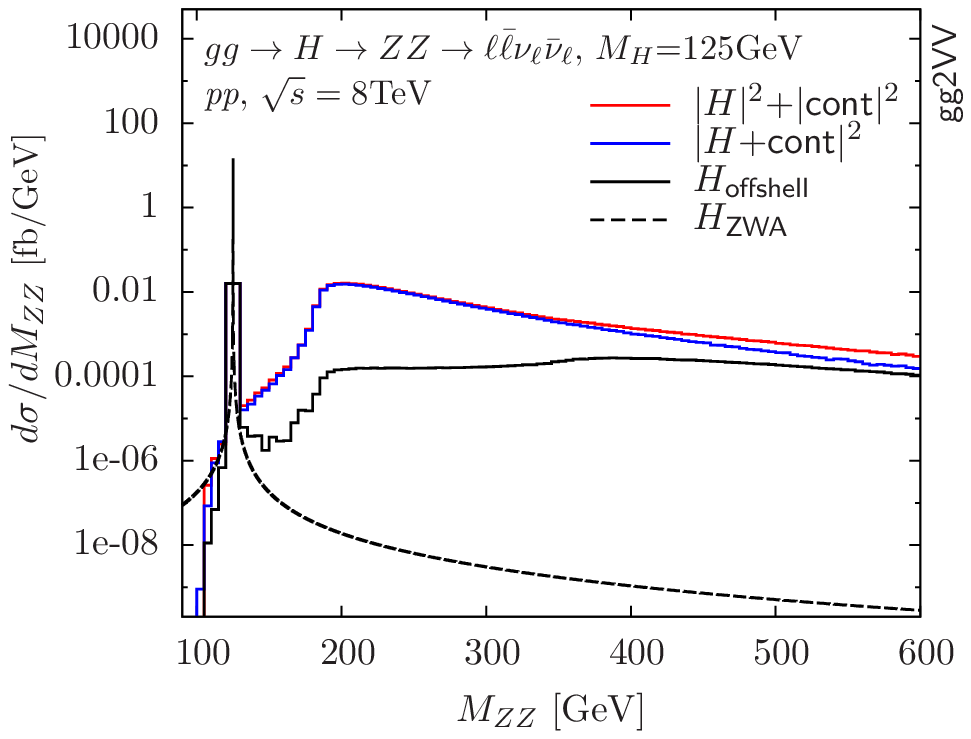}\\[.0cm]
\caption{\label{fig:2l2v_125_MZZ_l}
$M_{ZZ}$ distributions for $gg\to H\to ZZ\to \ell\bar{\ell}\nu_\ell\bar{\nu}_\ell$ 
for $M_H=125$\,GeV.
Applied cuts: $p_{T\ell} > 20$\,GeV, $|\eta_\ell| < 2.5$, 
$76$\,GeV $< M_{\ell\ell} < 106$\,GeV,
$\sla{p}_T > 10$\,GeV.
Other details as in Fig.\ \protect\ref{fig:lvlv_MWW_l}.
}
\end{figure}

\begin{figure}[t]
\vspace{0.4cm}
\centering
\includegraphics[width=0.7\textwidth, clip=true,angle=0]{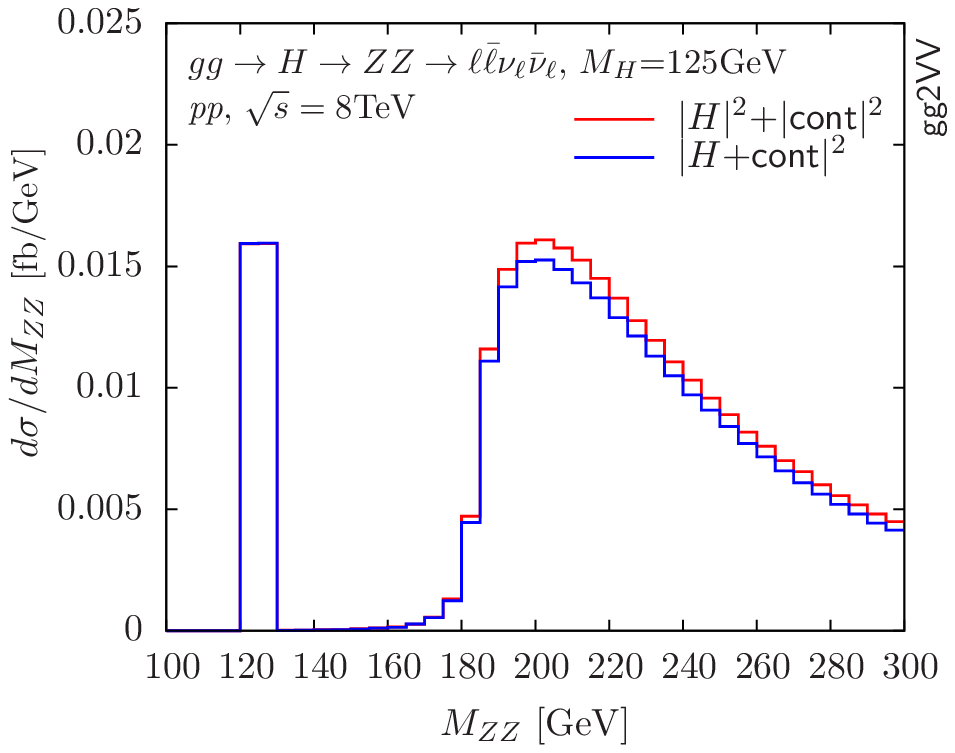}\\[.0cm]
\caption{\label{fig:2l2v_125_MZZ_m}
$M_{ZZ}$ distributions for $gg\to H\to ZZ\to \ell\bar{\ell}\nu_\ell\bar{\nu}_\ell$ 
for $M_H=125$\,GeV.
Interference effects in the region of the
Higgs resonance and the $Z$-pair threshold are shown.
Details as in Fig.\ \protect\ref{fig:2l2v_125_MZZ_l}.
}
\end{figure}

\begin{figure}[t]
\vspace{0.4cm}
\centering
\includegraphics[width=0.7\textwidth, clip=true,angle=0]{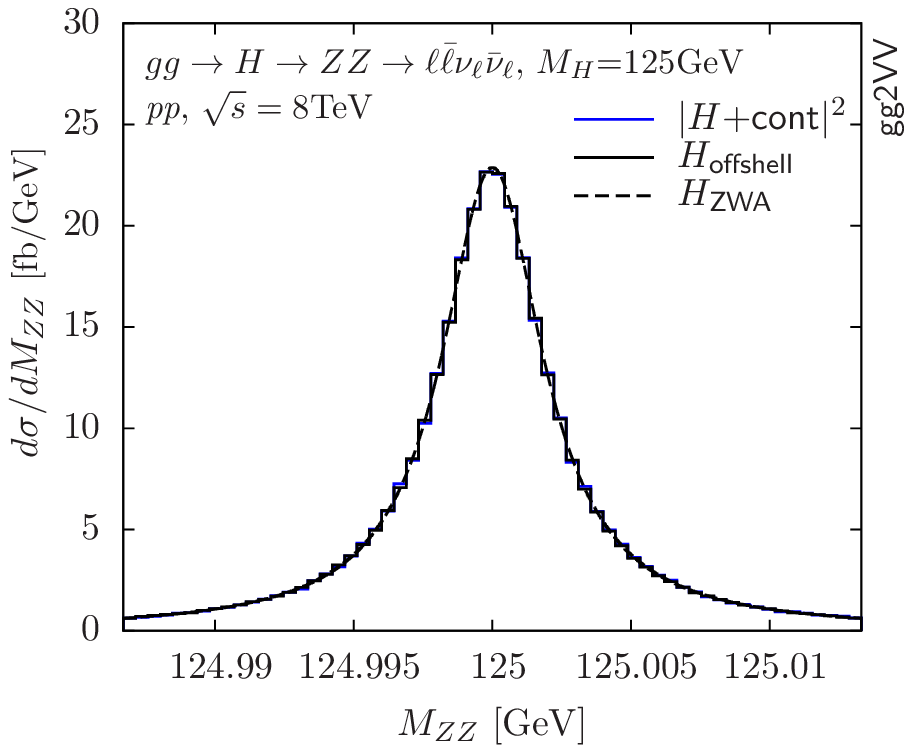}\\[.0cm]
\caption{\label{fig:2l2v_125_MZZ_s}
$M_{ZZ}$ distributions for $gg\to H\to ZZ\to \ell\bar{\ell}\nu_\ell\bar{\nu}_\ell$ 
for $M_H=125$\,GeV.
Off-shell and interference effects in the vicinity of the
Higgs resonance are shown.
Details as in Fig.\ \protect\ref{fig:2l2v_125_MZZ_l}.
}
\end{figure}

Given the rapid increase in integrated luminosity at the LHC, the 
$ZZ\to2\ell\,2\nu$ mode could also be of interest at $M_H=125$\,GeV.
We therefore extend our study to this Higgs mass. The following selection cuts
are applied: 
$p_{T\ell} > 20$\,GeV, $|\eta_\ell| < 2.5$, $76$\,GeV $< M_{\ell\ell} < 106$\,GeV
and $\sla{p}_T > 10$\,GeV.
As seen in Fig.\ \ref{fig:2l2v_125_MZZ_l}, the off-shell enhancement of the 
high $M_{H^\ast}$ tail is particularly pronounced.
In Table \ref{tab:2l2v_125}, cross section results are given.
The phase space region where $M_{ZZ}>180$\,GeV,
or equivalently $M_{ZZ} > M_H + 12000\Gamma_H$,
contributes 37\% to the off-shell Higgs cross section.  
The ZWA underestimates the Higgs cross section by a similar
amount.  
Figs.\ \ref{fig:2l2v_125_MZZ_l} and \ref{fig:2l2v_125_MZZ_m} illustrate that 
the region with $M_{ZZ} > 2\,M_Z$ is also responsible for interference effects
of $\calO(10\%)$.
Fig.\ \ref{fig:2l2v_125_MZZ_s} demonstrates that finite-width effects and
Higgs-continuum interference are negligible in the resonance region.

\begin{figure}[t]
\vspace{0.4cm}
\centering
\includegraphics[width=0.7\textwidth, clip=true,angle=0]{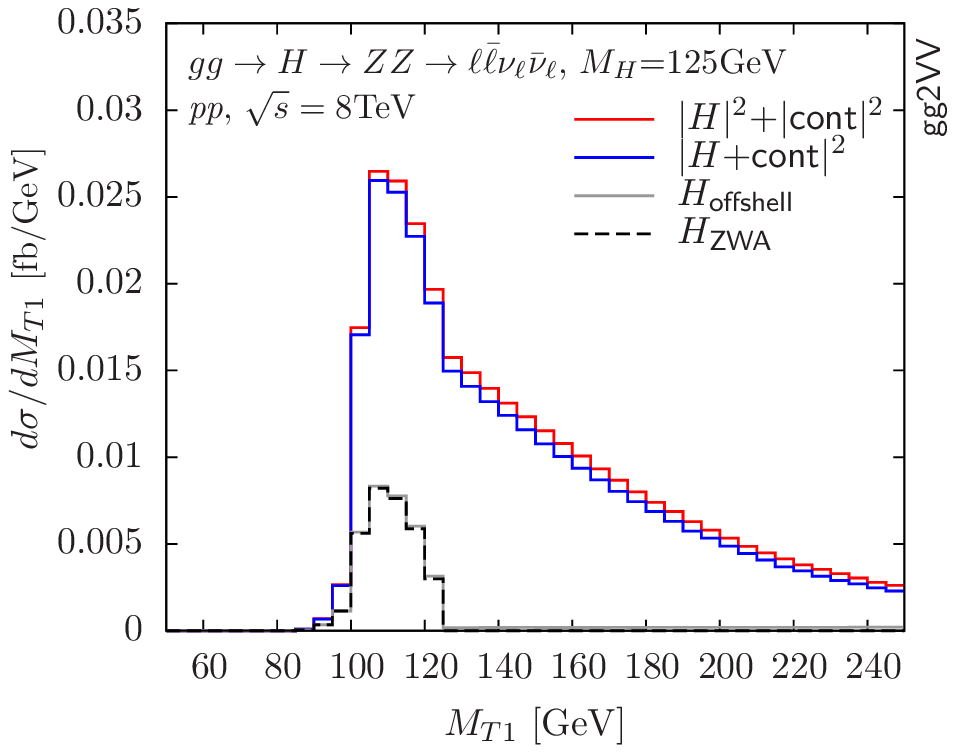}\\[.0cm]
\caption{\label{fig:2l2v_125_MT1}
Transverse mass distributions for 
$gg\to H\to ZZ\to \ell\bar{\ell}\nu_\ell\bar{\nu}_\ell$ 
for $M_H=125$\,GeV.
$M_{T1}$ is defined in Eq.\ (\ref{eq:MT1}) in the main text.
Other details as in Fig.\ \protect\ref{fig:2l2v_125_MZZ_l}.
}
\end{figure}

To mitigate the impact of the $M_{H^\ast}$ region with large ZWA deviations and 
Higgs-continuum interference, we propose to employ a $M_{T1}< M_H$ cut.
With this cut, the off-shell and interference effects ($R_1$) are
reduced to the 2\% level.  The $M_{T1}$ distribution displayed in
Fig.\ \ref{fig:2l2v_125_MT1} shows that the contamination of the $M_{T1}< M_H$
region from the interference-inducing $M_{H^\ast}>2\,M_Z$ region
is more severe than in the $WW$ case (see Fig.\ \ref{fig:lvlv_MT1_m}).

\begin{table}[t]
\vspace{0.4cm}
\renewcommand{\arraystretch}{1.2}
\centering
{\footnotesize
\begin{tabular}{|c|cccc|c|cc|}
\cline{2-5} 
\multicolumn{1}{c|}{} & \multicolumn{4}{|c|}{$gg\ (\to H)\to ZZ \to \ell\bar{\ell}\nu_\ell\bar{\nu}_\ell$} & \multicolumn{2}{|c}{} \\
\cline{2-8} 
\multicolumn{1}{c|}{} & \multicolumn{4}{|c|}{$\sigma$ [fb], $pp$, $\sqrt{s} = 8$\,TeV, $M_H=125$\,GeV} & ZWA & \multicolumn{2}{c|}{interference} \\
\hline
\multicolumn{1}{|c|}{$M_T$ cut} & $H_\text{ZWA}$ & $H_\text{offshell}$ & cont & $|H_\text{ofs}$+cont$|^2$ & $R_0$ & $R_1$ & $R_2$ \\
\hline
none & 0.1593(2) & 0.2571(2) & 1.5631(7) & 1.6376(9) & 0.6196(7) & 0.8997(6) & 0.290(5) \\
$M_{T1} < M_H$ & 0.1593(2) & 0.1625(2) & 0.4197(5) & 0.5663(6) & 0.980(2) & 0.973(2) & 0.902(5) \\
\hline
\end{tabular}}\\[.0cm]
\caption{\label{tab:2l2v_125}
Cross sections for $gg\ (\to H)\to ZZ \to \ell\bar{\ell}\nu_\ell\bar{\nu}_\ell$ for $M_H=125$\,GeV without and with transverse mass cut.
Applied cuts: $p_{T\ell} > 20$\,GeV, $|\eta_\ell| < 2.5$, 
$76$\,GeV $< M_{\ell\ell} < 106$\,GeV,
$\sla{p}_T > 10$\,GeV.
Other details as in Table \protect\ref{tab:2l2l_4l}.
}
\end{table}

\section{Conclusions \label{sec:conclusions}}
In the Higgs search at the LHC, a light Higgs boson is not 
excluded by experimental data.  
In the mass range $115$\,GeV $\lesssim M_H \lesssim$ 130\,GeV, one has
$\Gamma_H/M_H< 10^{-4}$ for the SM Higgs boson.
We have shown for inclusive cross sections and cross sections with 
experimental selection cuts that the ZWA is in general
not adequate and the error estimate $\calO(\Gamma_H/M_H)$
is not reliable for a light Higgs boson.  
The inclusion of off-shell 
contributions is essential to obtain an accurate Higgs signal 
normalisation at the 1\% precision level.  
We have traced this back to the dependence of the decay (and to a lesser degree 
production) matrix 
element on the Higgs virtuality $q^2$.  For the $H\to WW,ZZ$ decay modes we find 
that above the weak-boson pair production threshold the $(q^2)^2$ dependence of the decay 
matrix element compensates the $q^2$-dependence of the Higgs propagator, which 
results in a significantly enhanced off-shell cross section in comparison
to the ZWA cross section, when this phase space region, which is also affected by 
sizable Higgs-continuum interference, contributes.
As shape of the enhancement above $2\,M_V$, we find a ``plateau''
up to the $t\bar{t}$-threshold and an exponential decrease beyond it.  
The total $gg\to H \to VV$ cross section thus receives an ${\cal O}(10\%)$ 
off-shell correction.  We have further illustrated that the region above $2\,M_V$
is responsible for ${\cal O}(10\%)$ Higgs-continuum interference effects, which, 
due to the off-shell enhanced tail, can have a significant impact even for 
$M_H\ll 2\,M_V$.  We find that in the vicinity of the Higgs resonance finite-width
and Higgs-continuum interference effects are negligible for $M_H\ll 2\,M_V$,
while for $M_H=200$ GeV this is not the case.

For weak boson decays that permit the reconstruction 
of the invariant Higgs mass, the enhanced region is eliminated
by the experimental procedure as long as $M_H\ll 2\,M_V$.
For channels where the Higgs invariant mass cannot be reconstructed,
we have illustrated that $H\to WW$ search selection cuts for a light
Higgs boson reduce the contribution of the off-shell enhanced tail, and that 
the tail can effectively be excluded for the $H\to WW$ as well as $H\to ZZ$ 
decay mode by using transverse mass observables ($M_T$)
that approximate the Higgs invariant mass and applying a $M_T<M_H$ cut.
We predict that the weak boson fusion $H\to VV$ channels also 
exhibit an off-shell enhanced tail, since the effect is primarily 
caused by the Higgs decay matrix element.
It is worth noting that we make no statement about the observability of 
a large invariant mass signal due to a low-mass Higgs boson, which is 
hampered by the huge background and interference effects.

After the $5\sigma$-observation of a SM-like Higgs signal at 
$M_H\approx 125$--$126$ GeV reported by 
ATLAS and CMS in a recent seminar, the next step in
the analysis will be the extraction of the Higgs couplings and properties.
This study will initially be performed using the ZWA with a consequent error of
${\cal O}(5\%)$ on the couplings.  Although this is still tolerable with
current statistics, the results presented above make it clear that off-shell 
effects have to be included in future analyses.

In summary, we have elucidated the inadequacy of the ZWA in general and the 
existence of an enhanced tail in the Higgs invariant mass distribution of a light 
Higgs boson that decays to a weak-boson pair in particular, which
makes off-shell calculations mandatory and will lead to significant errors 
when the ZWA is used, unless the affected region of invariant masses above
the weak-boson-pair threshold is excluded with selection cuts.  The latter 
is also motivated by the fact that intuitively one would not like to assign 
events with large invariant mass to a low-mass Higgs signal, not least 
because they are affected by large signal-background interference.  As 
consequence of our findings, we recommend that the explicit or implicit 
application of the ZWA in experimental studies be identified and corrected 
for.


\acknowledgments

We would like to thank the conveners and 
members of the LHC Higgs Cross Section Working Group 
for stimulating and informative discussions.  N.K.\ would 
like to thank M.\ Rodgers for useful comparisons.
G.P.'s work is supported by MIUR under contract
2001023713$\_$006 and by Compagnia di San Paolo under contract ORTO11TPXK.
Financial support from the Higher Education Funding Council for England, 
the Science and Technology Facilities Council and the Institute for 
Particle Physics Phenomenology, Durham, is gratefully acknowledged
by N.K.




\begin{thebibliography}{22}


\bibitem{Higgs:1964ia}
  P.~W.~Higgs,
  \emph{Broken symmetries, massless particles and gauge fields},
  Phys.\ Lett.\  {\bf 12} (1964) 132.

\bibitem{Higgs:1964pj}
  P.~W.~Higgs,
  \emph{Broken symmetries and the masses of gauge bosons},
  Phys.\ Rev.\ Lett.\  {\bf 13} (1964) 508.

\bibitem{Higgs:1966ev}
  P.~W.~Higgs,
  \emph{Spontaneous symmetry breakdown without massless bosons},
  Phys.\ Rev.\  {\bf 145} (1966) 1156.

\bibitem{Englert:1964et}
  F.~Englert and R.~Brout,
  \emph{Broken symmetry and the mass of gauge vector mesons},
  Phys.\ Rev.\ Lett.\  {\bf 13} (1964) 321.

\bibitem{Guralnik:1964eu}
  G.~S.~Guralnik, C.~R.~Hagen and T.~W.~B.~Kibble,
  \emph{Global conservation laws and massless particles},
  Phys.\ Rev.\ Lett.\  {\bf 13} (1964) 585.


\bibitem{Barate:2003sz}
  R.~Barate {\it et al.}  [LEP Working Group for Higgs boson searches and ALEPH and DELPHI and L3 and OPAL Collaborations],
  \emph{Search for the Standard Model Higgs boson at LEP},
  Phys.\ Lett.\ B {\bf 565} (2003) 61
  [hep-ex/0306033].

\bibitem{:2012zzl}
  C.~a.~D.~C.~a.~t.~T.~N.~P.~a.~H.~W.~Group [Tevatron New Physics Higgs Working Group and CDF and D0 Collaborations],
  \emph{Updated combination of CDF and D0 searches for Standard Model Higgs boson production with up to 10.0 fb$^{-1}$ of data},
  arXiv:1207.0449 [hep-ex].

\bibitem{Chatrchyan:2012tx}
  S.~Chatrchyan {\it et al.}  [CMS Collaboration],
  \emph{Combined results of searches for the Standard Model Higgs boson in $pp$ collisions at $\sqrt{s} = 7$ TeV},
  Phys.\ Lett.\ B {\bf 710} (2012) 26
  [arXiv:1202.1488 [hep-ex]].

\bibitem{:2012an}
  G.~Aad {\it et al.}  [ATLAS Collaboration],
  \emph{Combined search for the Standard Model Higgs boson in $pp$ collisions at $\sqrt{s} = 7$ TeV with the ATLAS detector},
  arXiv:1207.0319 [hep-ex].


\bibitem{Georgi:1977gs}
  H.~M.~Georgi, S.~L.~Glashow, M.~E.~Machacek and D.~V.~Nanopoulos,
  \emph{Higgs bosons from two gluon annihilation in proton proton collisions},
  Phys.\ Rev.\ Lett.\  {\bf 40} (1978) 692.

\bibitem{Dawson:1990zj}
  S.~Dawson,
  \emph{Radiative corrections to Higgs boson production},
  Nucl.\ Phys.\ B {\bf 359} (1991) 283.

\bibitem{Djouadi:1991tka}
  A.~Djouadi, M.~Spira and P.~M.~Zerwas,
  \emph{Production of Higgs bosons in proton colliders: QCD corrections},
  Phys.\ Lett.\ B {\bf 264} (1991) 440.

\bibitem{Graudenz:1992pv}
  D.~Graudenz, M.~Spira and P.~M.~Zerwas,
  \emph{QCD corrections to Higgs boson production at proton proton colliders},
  Phys.\ Rev.\ Lett.\  {\bf 70} (1993) 1372.

\bibitem{Spira:1995rr}
  M.~Spira, A.~Djouadi, D.~Graudenz and P.~M.~Zerwas,
  \emph{Higgs boson production at the LHC},
  Nucl.\ Phys.\ B {\bf 453} (1995) 17
  [hep-ph/9504378].

\bibitem{Harlander:2002wh}
  R.~V.~Harlander and W.~B.~Kilgore,
  \emph{Next-to-next-to-leading order Higgs production at hadron colliders},
  Phys.\ Rev.\ Lett.\  {\bf 88} (2002) 201801
  [hep-ph/0201206].

\bibitem{Anastasiou:2002yz}
  C.~Anastasiou and K.~Melnikov,
  \emph{Higgs boson production at hadron colliders in NNLO QCD},
  Nucl.\ Phys.\  B {\bf 646} (2002) 220
  [arXiv:hep-ph/0207004].

\bibitem{Ravindran:2003um}
  V.~Ravindran, J.~Smith and W.~L.~van Neerven,
  \emph{NNLO corrections to the total cross section for Higgs boson production in
  hadron-hadron collisions},
  Nucl.\ Phys.\  B {\bf 665} (2003) 325
  [arXiv:hep-ph/0302135].


\bibitem{Catani:2003zt}
  S.~Catani, D.~de Florian, M.~Grazzini and P.~Nason,
  \emph{Soft gluon resummation for Higgs boson production at hadron colliders},
  JHEP {\bf 0307} (2003) 028
  [hep-ph/0306211].

\bibitem{deFlorian:2011xf}
  D.~de Florian, G.~Ferrera, M.~Grazzini and D.~Tommasini,
  \emph{Transverse-momentum resummation: Higgs boson production at the Tevatron and the LHC},
  JHEP {\bf 1111} (2011) 064
  [arXiv:1109.2109 [hep-ph]].

\bibitem{Moch:2005ky}
  S.~Moch and A.~Vogt,
  \emph{Higher-order soft corrections to lepton pair and Higgs boson production},
  Phys.\ Lett.\ B {\bf 631} (2005) 48
  [hep-ph/0508265].

\bibitem{Laenen:2005uz}
  E.~Laenen and L.~Magnea,
  \emph{Threshold resummation for electroweak annihilation from DIS data},
  Phys.\ Lett.\ B {\bf 632} (2006) 270
  [hep-ph/0508284].

\bibitem{Idilbi:2005ni}
  A.~Idilbi, X.~-d.~Ji, J.~-P.~Ma and F.~Yuan,
  \emph{Threshold resummation for Higgs production in effective field theory},
  Phys.\ Rev.\ D {\bf 73} (2006) 077501
  [hep-ph/0509294].

\bibitem{Ravindran:2005vv}
  V.~Ravindran,
  \emph{On Sudakov and soft resummations in QCD},
  Nucl.\ Phys.\ B {\bf 746} (2006) 58
  [hep-ph/0512249].

\bibitem{Ravindran:2006cg}
  V.~Ravindran,
  \emph{Higher-order threshold effects to inclusive processes in QCD},
  Nucl.\ Phys.\ B {\bf 752} (2006) 173
  [hep-ph/0603041].

\bibitem{Ahrens:2008nc}
  V.~Ahrens, T.~Becher, M.~Neubert and L.~L.~Yang,
  \emph{Renormalization-group improved prediction for Higgs production at hadron colliders},
  Eur.\ Phys.\ J.\ C {\bf 62} (2009) 333
  [arXiv:0809.4283 [hep-ph]].


\bibitem{Anastasiou:2004xq}
  C.~Anastasiou, K.~Melnikov and F.~Petriello,
  \emph{Higgs boson production at hadron colliders: differential cross sections
  through next-to-next-to-leading order},
  Phys.\ Rev.\ Lett.\  {\bf 93} (2004) 262002
  [arXiv:hep-ph/0409088].

\bibitem{Catani:2007vq}
  S.~Catani and M.~Grazzini,
  \emph{A NNLO subtraction formalism in hadron collisions and its application to
  Higgs boson production at the LHC},
  Phys.\ Rev.\ Lett.\  {\bf 98} (2007) 222002
  [arXiv:hep-ph/0703012].


\bibitem{Marzani:2008az}
  S.~Marzani, R.~D.~Ball, V.~Del Duca, S.~Forte and A.~Vicini,
  \emph{Higgs production via gluon-gluon fusion with finite top mass beyond next-to-leading order},
  Nucl.\ Phys.\ B {\bf 800} (2008) 127
  [arXiv:0801.2544 [hep-ph]].

\bibitem{Harlander:2009bw}
  R.~V.~Harlander and K.~J.~Ozeren,
  \emph{Top mass effects in Higgs production at next-to-next-to-leading order QCD: Virtual corrections},
  Phys.\ Lett.\ B {\bf 679} (2009) 467
  [arXiv:0907.2997 [hep-ph]].

\bibitem{Pak:2009bx}
  A.~Pak, M.~Rogal and M.~Steinhauser,
  \emph{Virtual three-loop corrections to Higgs boson production in gluon fusion for finite top quark mass},
  Phys.\ Lett.\ B {\bf 679} (2009) 473
  [arXiv:0907.2998 [hep-ph]].

\bibitem{Harlander:2009mq}
  R.~V.~Harlander and K.~J.~Ozeren,
  \emph{Finite top mass effects for hadronic Higgs production at next-to-next-to-leading order},
  JHEP {\bf 0911} (2009) 088
  [arXiv:0909.3420 [hep-ph]].

\bibitem{Pak:2009dg}
  A.~Pak, M.~Rogal and M.~Steinhauser,
  \emph{Finite top quark mass effects in NNLO Higgs boson production at LHC},
  JHEP {\bf 1002} (2010) 025
  [arXiv:0911.4662 [hep-ph]].

\bibitem{Harlander:2009my}
  R.~V.~Harlander, H.~Mantler, S.~Marzani and K.~J.~Ozeren,
  \emph{Higgs production in gluon fusion at next-to-next-to-leading order QCD for finite top mass},
  Eur.\ Phys.\ J.\ C {\bf 66} (2010) 359
  [arXiv:0912.2104 [hep-ph]].

\bibitem{Baglio:2010ae}
  J.~Baglio and A.~Djouadi,
  \emph{Higgs production at the lHC},
  JHEP {\bf 1103} (2011) 055
  [arXiv:1012.0530 [hep-ph]].



\bibitem{Djouadi:1994ge}
  A.~Djouadi and P.~Gambino,
  \emph{Leading electroweak correction to Higgs boson production at proton colliders},
  Phys.\ Rev.\ Lett.\  {\bf 73} (1994) 2528
  [hep-ph/9406432].

\bibitem{Aglietti:2004nj}
  U.~Aglietti, R.~Bonciani, G.~Degrassi and A.~Vicini,
  \emph{Two loop light fermion contribution to Higgs production and decays},
  Phys.\ Lett.\ B {\bf 595} (2004) 432
  [hep-ph/0404071].

\bibitem{Degrassi:2004mx}
  G.~Degrassi and F.~Maltoni,
  \emph{Two-loop electroweak corrections to Higgs production at hadron colliders},
  Phys.\ Lett.\ B {\bf 600} (2004) 255
  [hep-ph/0407249].

\bibitem{Actis:2008ug}
  S.~Actis, G.~Passarino, C.~Sturm and S.~Uccirati,
  \emph{NLO electroweak corrections to Higgs boson production at hadron colliders},
  Phys.\ Lett.\ B {\bf 670} (2008) 12
  [arXiv:0809.1301 [hep-ph]].

\bibitem{Actis:2008ts}
  S.~Actis, G.~Passarino, C.~Sturm and S.~Uccirati,
  \emph{NNLO computational techniques: the cases $H\to\gamma\gamma$ and $H\to gg$},
  Nucl.\ Phys.\ B {\bf 811} (2009) 182
  [arXiv:0809.3667 [hep-ph]].

\bibitem{Ahrens:2010rs}
  V.~Ahrens, T.~Becher, M.~Neubert and L.~L.~Yang,
  \emph{Updated predictions for Higgs production at the Tevatron and the LHC},
  Phys.\ Lett.\ B {\bf 698} (2011) 271
  [arXiv:1008.3162 [hep-ph]].

\bibitem{Keung:2009bs}
  W.~-Y.~Keung and F.~J.~Petriello,
  \emph{Electroweak and finite quark-mass effects on the Higgs boson transverse momentum distribution},
  Phys.\ Rev.\ D {\bf 80} (2009) 013007
  [arXiv:0905.2775 [hep-ph]].

\bibitem{Brein:2010xj}
  O.~Brein,
  \emph{Electroweak and bottom quark contributions to Higgs boson plus jet production},
  Phys.\ Rev.\ D {\bf 81} (2010) 093006
  [arXiv:1003.4438 [hep-ph]].

\bibitem{Anastasiou:2008tj}
  C.~Anastasiou, R.~Boughezal and F.~Petriello,
  \emph{Mixed QCD-electroweak corrections to Higgs boson production in gluon fusion},
  JHEP {\bf 0904} (2009) 003
  [arXiv:0811.3458 [hep-ph]].


\bibitem{deFlorian:2009hc}
  D.~de Florian and M.~Grazzini,
  \emph{Higgs production through gluon fusion: Updated cross sections at the Tevatron and the LHC},
  Phys.\ Lett.\ B {\bf 674} (2009) 291
  [arXiv:0901.2427 [hep-ph]].

\bibitem{Dittmaier:2011ti}
  S.~Dittmaier {\it et al.}  [LHC Higgs Cross Section Working Group Collaboration],
  \emph{Handbook of LHC Higgs Cross Sections: 1.\ Inclusive observables},
  arXiv:1101.0593 [hep-ph].

\bibitem{Anastasiou:2011pi}
  C.~Anastasiou, S.~Buehler, F.~Herzog and A.~Lazopoulos,
  \emph{Total cross section for Higgs boson hadroproduction with anomalous Standard Model interactions},
  JHEP {\bf 1112} (2011) 058
  [arXiv:1107.0683 [hep-ph]].

\bibitem{Anastasiou:2012hx}
  C.~Anastasiou, S.~Buehler, F.~Herzog and A.~Lazopoulos,
  \emph{Inclusive Higgs boson cross-section for the LHC at 8 TeV},
  JHEP {\bf 1204} (2012) 004
  [arXiv:1202.3638 [hep-ph]].

\bibitem{deFlorian:2012yg}
  D.~de Florian and M.~Grazzini,
  \emph{Higgs production at the LHC: updated cross sections at $\sqrt{s}=8$ TeV},
  arXiv:1206.4133 [hep-ph].


\bibitem{Anastasiou:2007mz}
  C.~Anastasiou, G.~Dissertori and F.~Stockli,
  \emph{NNLO QCD predictions for the $H\to WW\to \ell\nu \ell\nu$ signal at the LHC},
  JHEP {\bf 0709} (2007) 018
  [arXiv:0707.2373 [hep-ph]].

\bibitem{Grazzini:2008tf}
  M.~Grazzini,
   \emph{NNLO predictions for the Higgs boson signal in the $H\to WW\to \ell\nu \ell\nu$ 
   and $H\to ZZ\to 4\ell$ decay channels},
  JHEP {\bf 0802} (2008) 043
  [arXiv:0801.3232 [hep-ph]].

\bibitem{Dittmaier:2012vm}
  S.~Dittmaier, C.~Mariotti, G.~Passarino, R.~Tanaka, S.~Alekhin, J.~Alwall and E.~A.~Bagnaschi {\it et al.},
  \emph{Handbook of LHC Higgs cross sections: 2.\ Differential distributions},
  arXiv:1201.3084 [hep-ph].


\bibitem{Bredenstein:2006rh}
  A.~Bredenstein, A.~Denner, S.~Dittmaier and M.~M.~Weber,
  \emph{Precise predictions for the Higgs-boson decay $H\to WW/ZZ \to$ 4 leptons},
  Phys.\ Rev.\ D {\bf 74} (2006) 013004
  [hep-ph/0604011].

\bibitem{Denner:2011vt}
  A.~Denner, S.~Dittmaier, A.~Muck, G.~Passarino, M.~Spira, C.~Sturm, S.~Uccirati and M.~M.~Weber,
  \emph{Higgs production and decay with a fourth Standard-Model-like fermion generation},
  Eur.\ Phys.\ J.\ C {\bf 72} (2012) 1992
  [arXiv:1111.6395 [hep-ph]].


\bibitem{Goria:2011wa}
  S.~Goria, G.~Passarino and D.~Rosco,
  \emph{The Higgs boson lineshape},
  arXiv:1112.5517 [hep-ph].

\bibitem{Buehler:2012zf}
  S.~Buehler,
  \emph{Precise inclusive Higgs predictions using iHixs},
  arXiv:1201.0985 [hep-ph].


\bibitem{Berdine:2007uv}
  D.~Berdine, N.~Kauer and D.~Rainwater,
  \emph{Breakdown of the narrow-width approximation for New Physics},
  Phys.\ Rev.\ Lett.\  {\bf 99} (2007) 111601
  [hep-ph/0703058].

\bibitem{Kauer:2007zc}
  N.~Kauer,
  \emph{Narrow-width approximation limitations},
  Phys.\ Lett.\ B {\bf 649} (2007) 413
  [hep-ph/0703077].

\bibitem{Kauer:2007nt}
  N.~Kauer,
  \emph{A threshold-improved narrow-width approximation for BSM physics},
  JHEP {\bf 0804} (2008) 055
  [arXiv:0708.1161 [hep-ph]].

\bibitem{Uhlemann:2008pm}
  C.~F.~Uhlemann and N.~Kauer,
  \emph{Narrow-width approximation accuracy},
  Nucl.\ Phys.\ B {\bf 814} (2009) 195
  [arXiv:0807.4112 [hep-ph]].


\bibitem{Glover:1988fe}
  E.~W.~N.~Glover and J.~J.~van der Bij,
  \emph{Vector boson pair production via gluon fusion},
  Phys.\ Lett.\ B {\bf 219} (1989) 488.

\bibitem{Glover:1988rg}
  E.~W.~N.~Glover and J.~J.~van der Bij,
  \emph{$Z$-boson pair production via gluon fusion},
  Nucl.\ Phys.\ B {\bf 321} (1989) 561.

\bibitem{Binoth:2006mf}
  T.~Binoth, M.~Ciccolini, N.~Kauer and M.~Kramer,
  \emph{Gluon-induced $W$-boson pair production at the LHC},
  JHEP {\bf 0612} (2006) 046
  [hep-ph/0611170].

\bibitem{Accomando:2007xc}
  E.~Accomando,
  \emph{The process $gg\to WW$ as a probe into the EWSB mechanism},
  Phys.\ Lett.\ B {\bf 661} (2008) 129
  [arXiv:0709.1364 [hep-ph]].

\bibitem{Campbell:2011cu}
  J.~M.~Campbell, R.~K.~Ellis and C.~Williams,
  \emph{Gluon-gluon contributions to $W^+W^-$ production and Higgs interference effects},
  JHEP {\bf 1110} (2011) 005
  [arXiv:1107.5569 [hep-ph]].

\bibitem{Kauer:2012ma}
  N.~Kauer,
  \emph{Signal-background interference in $gg \to H \to VV$},
  arXiv:1201.1667 [hep-ph].

\bibitem{Dixon:2003yb}
  L.~J.~Dixon and M.~S.~Siu,
  \emph{Resonance-continuum interference in the diphoton Higgs signal at the LHC},
  Phys.\ Rev.\ Lett.\  {\bf 90} (2003) 252001
  [hep-ph/0302233].

\bibitem{Dixon:2008xc}
  L.~J.~Dixon and Y.~Sofianatos,
  \emph{Resonance-continuum interference in light Higgs boson production at a photon collider},
  Phys.\ Rev.\ D {\bf 79} (2009) 033002
  [arXiv:0812.3712 [hep-ph]].

\bibitem{Accomando:2011eu}
  E.~Accomando, D.~Becciolini, S.~De Curtis, D.~Dominici, L.~Fedeli and C.~Shepherd-Themistocleous,
  \emph{Interference effects in heavy $W'$-boson searches at the LHC},
  arXiv:1110.0713 [hep-ph].


\bibitem{Campbell:2011bn}
  J.~M.~Campbell, R.~K.~Ellis and C.~Williams,
  \emph{Vector boson pair production at the LHC},
  JHEP {\bf 1107} (2011) 018
  [arXiv:1105.0020 [hep-ph]].

\bibitem{Frederix:2011ss}
  R.~Frederix, S.~Frixione, V.~Hirschi, F.~Maltoni, R.~Pittau and P.~Torrielli,
  \emph{Four-lepton production at hadron colliders: aMC@NLO predictions with theoretical uncertainties},
  JHEP {\bf 1202} (2012) 099
  [arXiv:1110.4738 [hep-ph]].

\bibitem{Melia:2012zg}
  T.~Melia, K.~Melnikov, R.~Rontsch, M.~Schulze and G.~Zanderighi,
  \emph{Gluon fusion contribution to $W^+W^-+\,$jet production},
  arXiv:1205.6987 [hep-ph].

\bibitem{Agrawal:2012df}
  P.~Agrawal and A.~Shivaji,
  \emph{Di-vector boson + jet production via gluon fusion at hadron colliders},
  arXiv:1207.2927 [hep-ph].


\bibitem{Kauer:2001sp}
  N.~Kauer and D.~Zeppenfeld,
  \emph{Finite-width effects in top quark production at hadron colliders},
  Phys.\ Rev.\ D {\bf 65} (2002) 014021
  [hep-ph/0107181].


\bibitem{Passarino:2010qk}
  G.~Passarino, C.~Sturm and S.~Uccirati,
  \emph{Higgs pseudo-observables, second Riemann sheet and all that},
  Nucl.\ Phys.\ B {\bf 834} (2010) 77
  [arXiv:1001.3360 [hep-ph]].

\bibitem{Actis:2006rc}
  S.~Actis and G.~Passarino,
  \emph{Two-loop renormalization in the Standard Model, Part III: Renormalization equations and their solutions},
  Nucl.\ Phys.\ B {\bf 777} (2007) 100
  [hep-ph/0612124].

\bibitem{Bredenstein:2007ec}
  A.~Bredenstein, A.~Denner, S.~Dittmaier and M.~M.~Weber,
  \emph{Precision calculations for $H\to WW/ZZ \to 4$ fermions with PROPHECY4f},
  arXiv:0708.4123 [hep-ph].

\bibitem{Martin:2009iq}
  A.~D.~Martin, W.~J.~Stirling, R.~S.~Thorne and G.~Watt,
  \emph{Parton distributions for the LHC},
  Eur.\ Phys.\ J.\ C {\bf 63} (2009) 189
  [arXiv:0901.0002 [hep-ph]].


\bibitem{deFlorianPrivComm}
D.~de~Florian (2012) private communication.

\bibitem{RebuzziPrivComm}
D.~Rebuzzi (2012) private communication.

\bibitem{VBFTwiki}
\url{https://twiki.cern.ch/twiki/bin/view/LHCPhysics/VBF}


\bibitem{HWWcuts}
\url{https://twiki.cern.ch/twiki/bin/view/LHCPhysics/WW}


\bibitem{HZZcuts}
\url{https://twiki.cern.ch/twiki/bin/view/LHCPhysics/ZZ}


\bibitem{gg2VV}
\url{http://gg2VV.hepforge.org/}

\bibitem{Binoth:2005ua}
  T.~Binoth, M.~Ciccolini, N.~Kauer and M.~Kramer,
  \emph{Gluon-induced $WW$ background to Higgs boson searches at the LHC},
  JHEP {\bf 0503} (2005) 065
  [hep-ph/0503094].

\bibitem{Binoth:2008pr} 
  T.~Binoth, N.~Kauer and P.~Mertsch,
  \emph{Gluon-induced QCD corrections to $pp \to ZZ \to \ell\bar{\ell}\ell'\bar{\ell'}$},
  arXiv:0807.0024 [hep-ph].

\bibitem{Hahn:1998yk}             
  T.~Hahn and M.~Perez-Victoria,
  \emph{Automatized one-loop calculations in four and $D$ dimensions},
  Comput.\ Phys.\ Commun.\  {\bf 118} (1999) 153
  [arXiv:hep-ph/9807565].

\bibitem{Hahn:2000kx}          
  T.~Hahn,
  \emph{Generating Feynman diagrams and amplitudes with FeynArts 3},
  Comput.\ Phys.\ Commun.\  {\bf 140} (2001) 418
  [arXiv:hep-ph/0012260].


\bibitem{Djouadi:1997yw}
  A.~Djouadi, J.~Kalinowski and M.~Spira,
  \emph{HDECAY: a program for Higgs boson decays in the Standard Model and its supersymmetric extension},
  Comput.\ Phys.\ Commun.\  {\bf 108} (1998) 56
  [hep-ph/9704448].


\bibitem{ATLAS:2012ac}
  G.~Aad {\it et al.}  [ATLAS Collaboration],
  \emph{Search for the Standard Model Higgs boson in the decay channel $H\to ZZ^\ast\to 4\ell$ with 4.8fb$^{-1}$ of $pp$ collision data at $\sqrt{s}=7$\,TeV with ATLAS},
  Phys.\ Lett.\ B {\bf 710} (2012) 383
  [arXiv:1202.1415 [hep-ex]].

\bibitem{Chatrchyan:2012dg}
  S.~Chatrchyan {\it et al.}  [CMS Collaboration],
  \emph{Search for the Standard Model Higgs boson in the decay channel $H\to ZZ \to 4$ leptons in $pp$ collisions at $\sqrt{s}= 7$\,TeV},
  Phys.\ Rev.\ Lett.\  {\bf 108} (2012) 111804
  [arXiv:1202.1997 [hep-ex]].


\bibitem{ATLAS_lvlv}
  G.~Aad {\it et al.}  [ATLAS Collaboration],
  \emph{Search for the Standard Model Higgs boson in the $H \to  WW^\ast \to  \ell \nu \ell \nu$ decay mode with 4.7fb$^{-1}$ of ATLAS data at $\sqrt{s}= 7$\,TeV},
  arXiv:1206.0756 [hep-ex].

\bibitem{Chatrchyan:2012ty}
  S.~Chatrchyan {\it et al.}  [CMS Collaboration],
  \emph{Search for the Standard Model Higgs boson decaying to a $W$ pair in the fully leptonic final state in $pp$ collisions at $\sqrt{s}= 7$\,TeV},
  Phys.\ Lett.\ B {\bf 710} (2012) 91
  [arXiv:1202.1489 [hep-ex]].


\bibitem{Barr:2009mx}
  A.~J.~Barr, B.~Gripaios and C.~G.~Lester,
  \emph{Measuring the Higgs boson mass in dileptonic $W$-boson decays at hadron colliders},
  JHEP {\bf 0907} (2009) 072
  [arXiv:0902.4864 [hep-ph]].


\bibitem{ATLAS_2l2v}
  G.~Aad {\it et al.}  [ATLAS Collaboration],
  \emph{Search for a Standard Model Higgs boson in the $H \to  ZZ \to  \ell\ell\nu\nu$ decay channel using 4.7 fb$^{-1}$ of $\sqrt{s}= 7$\,TeV data with the ATLAS detector},
  arXiv:1205.6744 [hep-ex].

\bibitem{Chatrchyan:2012ft}
  S.~Chatrchyan {\it et al.}  [CMS Collaboration],
  \emph{Search for the Standard Model Higgs boson in the $H \to ZZ \to 2\ell 2\nu$ channel in $pp$ collisions at $\sqrt{s}= 7$\,TeV},
  JHEP {\bf 1203} (2012) 040
  [arXiv:1202.3478 [hep-ex]].


\bibitem{Rainwater:1999sd}
  D.~L.~Rainwater and D.~Zeppenfeld,
  \emph{Observing $H \to  W^\ast W^\ast \to  e^\pm \mu^\mp \sla{p}_T$ in weak boson fusion with dual forward jet tagging at the CERN LHC},
  Phys.\ Rev.\ D {\bf 60} (1999) 113004
   [Erratum-ibid.\ D {\bf 61} (2000) 099901]
  [hep-ph/9906218].


\end{thebibliography}
\end{document}